\documentclass[%
 aip,
amsmath,amssymb,
reprint, twocolumn 
]{revtex4-1}

\usepackage[version=3]{mhchem} 
\usepackage{mathtools}
\usepackage{gensymb}
\usepackage{multirow}
\usepackage{listings}

\usepackage[utf8]{inputenc}
\usepackage[T1]{fontenc}
\usepackage{mathptmx}
\usepackage{etoolbox}

\makeatletter
\def\@email#1#2{%
 \endgroup
 \patchcmd{\titleblock@produce}
  {\frontmatter@RRAPformat}
  {\frontmatter@RRAPformat{\produce@RRAP{*#1\href{mailto:#2}{#2}}}\frontmatter@RRAPformat}
  {}{}
}%
\makeatother
\begin{document}

\title {Open-source implementation of the anti-Hermitian contracted Schr\"odinger equation for electronic ground and excited states}

\author{Daniel Gibney}
\author{Anthony W. Schlimgen}
\author{Jan-Niklas Boyn}
\email{jboyn@umn.edu}
\email{aws@umn.edu}

\affiliation{Department of Chemistry, University of Minnesota, Minneapolis, Minnesota 55455, United States}

\date{\today}

\begin{abstract}
Efficient simulation of strongly correlated electrons has become a routine tool in molecular electronic structure theory due to recent advances in approximate configuration interaction (CI) techniques. Nonetheless, the quantitative and predictive description of molecular electronic states remains a significant challenge due to the difficulty of computing all-electron correlation beyond CI. Here, we describe a new open-source implementation of the anti-Hermitian contracted Schr\"odinger equation (ACSE) for use in accurate simulation of all-electron correlation in molecules. In contrast to standard approaches via multireference perturbation theory, the scaling of the ACSE does not depend on the complexity of the strongly correlated reference wavefunction. Furthermore, the ACSE employs the exact electronic Hamiltonian, rather than an approximate perturbative Hamiltonian. Our benchmark results demonstrate good accuracy for main group and transition metal systems, in weakly and strongly correlated regimes, with various basis sets, and for ground and excited states. The results suggest that the ACSE has potential as a scalable and robust technique for simulating all-electron correlation in molecular ground and excited states. 

\end{abstract}

\maketitle

\section{Introduction}

Strongly correlated electrons in molecules and materials present significant challenges to modern quantum chemistry. Cutting edge experimental work in catalysis, energy transformation and storage, and light-matter interaction frequently leverages the complexity of electronic wavefunctions to affect efficient physical transformations. Examples include high-spin, highly entangled multi-metallic transition metal catalysts,~\cite{Roelfes:2010,Ward:2018,Ward:2019,Karlin:1993,Nam:2007,Ribbe:2009} excited state decay in quantum dynamics,~\cite{Marian:2012aa,Curchod:2018aa} and applications of molecular magnetism.~\cite{Singh:2018aa,Singh:2020aa,Averkiev:2011aa} However, the \emph{ab initio} simulation of the electronic structure of these systems is difficult due to their multireference (MR) nature, meaning their wavefunctions require more than one Slater determinant for a good zeroth order approximation. 

Modern multireference theories have made great strides in the practical calculation of these complex chemical systems, with several popular approximations having been developed to overcome the exponential scaling limitations of traditional full configuration interaction (FCI) solvers. These include, for example, density matrix renormalization group (DMRG),\cite{Richard:1999,Chan:2011,Knecht:2016} stochastic and selected configuration interaction (CI),\cite{Smith:2017,Li:2018,Holmes:2016} as well as variational two-electron reduced density matrix (2-RDM) techniques.\cite{Mazziotti:2011,,Mazziotti:2012,Mostafanejad:2019} These techniques have expanded the feasible size of near-exact CI calculations, allowing for the routine treatment of approximate wavefunctions or density matrices comprised of dozens of electrons and orbitals. Approximate CI solvers provide a high-quality treatment of the \emph{strong} correlation in many-electron systems, which arises when multiple reference determinants have important contributions to the wavefunction. 

Even with the advances of large-scale CI solvers, approximations are still routinely required when simulating complex chemical systems due to the rapidly growing number of orbitals required for quantitative simulation. In practice, even large-scale approximate CI is often limited to the valence and semi-core orbitals and electrons, which is known as the \emph{active space}. In this context, the strong correlation described by the active-space CI must be supplemented by the \emph{dynamical} correlation outside of the active space. The standard approaches for all-electron correlation begin with a complete active space self-consistent field (CASSCF) calculation which provides the reference wavefunction for the dynamic correlation correction.~\cite{Siegbahn:1981,Roos:1987,Lischka:2018} Three particularly powerful approaches to dynamic correlation are MR perturbation theory (MRPT),~\cite{Andersson:1990aa,Angeli:2002aa,Guo:2016ab,Andersson:1992aa}  MR coupled cluster theory (MRCC),~\cite{Lyakh:2012aa} and the recently developed multi-configuration pair-density functional theory (MC-PDFT).~\cite{Mostafanejad:2020,Gagliardi:2017,Sharma:2021,Zhou:2022,Bao:2025,Feng:2025,Manni:2014,Hennefarth:2025}

An alternative to these approaches is the anti-Hermitian contracted Schr\"odinger equation (ACSE), which is related to the variance or dispersion relation for the Schr\"odinger equation.~\cite{Mazziotti:2006aa,Mazziotti:2007aa,Gidofalvi:2009aa} Formally the ACSE is expressed in terms of the 3-RDM; however, in practice, approximations relying only on the 2-RDM are employed, which allows for iterative minimization of a residual function. The ACSE treats the entire molecular orbital space on equal footing, so it can consistently describe all-electron correlation even for qualitatively different reference wavefunctions.~\cite{Mazziotti:2007ab,Mazziotti:2007ac} The ACSE also has the advantage of using the exact molecular Hamiltonian instead of approximate Hamiltonians, common in MRPT, which introduce intruder states or discontinuities in the potential surface.~\cite{Evangelisti:1987aa,vanvoorhis:2025}

Here, we present a new, open-source, python based, implementation of the ACSE for computing all-electron correlation from a reference wavefunction obtained from a Hartree-Fock or CASSCF calculation. As the ACSE minimizes the residual and not the energy, excited states may be targeted by using an excited state CASSCF reference wavefunction. Our implementation of the ACSE minimizes the residual equation on the space of 2-RDMs by exploiting approximate reconstructions of the 3-RDM. For a system with $r$ molecular orbitals, the resulting algorithm produces the approximate 2-RDM with $\mathcal{O}(r^6)$ computational scaling and $\mathcal{O}(r^4)$ memory requirements. Previous work has reported ACSE calculations of some small molecules and atoms;~\cite{Boyn:2022aa,Foley:2021aa,Gidofalvi:2009aa,Snyder:2010aa,Greenman:2010aa,Snyder:2011aa,Foley:2009aa,Greenman:2011aa,Schlimgen:2017aa,Mazziotti:2008aa} however, the broad use of this technique remains limited, owing to a lack of available, open-source codes, and systematic benchmarking. Here, we benchmark our implementation using several prototypical molecular systems including the atomization of linear H$_6$, the rotational barrier of ethylene, a vertical excitation of ethylene, the low energy spectrum of N$_2$ along its dissociation and the spin splitting energies of Fe$^{2+}$, Fe$^{3+}$, and Co$^{3+}$. Our results are a first step toward providing a comprehensive evaluation of the ACSE in larger basis sets and in varied molecular motifs.

\section{Theory}

The non-relativistic electronic Hamiltonian is a sum of one- and two-electron terms,
\begin{equation}
    H = \sum_{pr}{}^1K^p_ra^{\dagger}_pa_r + \sum_{pqrs}{}^2V^{pq}_{rs}a^{\dagger}_pa^{\dagger}_qa_sa_r \,,
\end{equation}
where $a^\dagger$ and $a$ are the fermionic creation and annihilation operators, respectively, and the matrices $^1K$ and $^2V$ contain the one- and two-electron integrals, respectively.~\cite{szabo:1989aa} The Hamiltonian can be expressed as a two-body operator,
\begin{equation}
    ^2K =\sum_{pqrs}{}^2V^{pq}_{rs}a^{\dagger}_pa^{\dagger}_qa_sa_r + \frac{1}{N-1}\bigg({}^1K^p_ra^{\dagger}_pa_r\delta_{qs}+{}^1K^q_sa^{\dagger}_qa_s\delta_{pr}\bigg),
\end{equation}
where $N$ is the number of electrons and $\delta$ is the Kronecker delta function.~\cite{Mazziotti:1998aa}

While there are many approaches to finding the eigenstates of this Hamiltonian, here we employ the contracted Schr\"odinger equation (CSE) formalism, which expresses the solution for the eigenstates in terms of a hierarchy of density matrices. Crucially, the solutions to the CSE have a one-to-one correspondence with the solutions to the SE, which was shown using the dispersion or variance in the SE.~\cite{Cohen:1976aa,Nakatsuji:1976aa,Harriman:1979aa,Mazziotti:1998aa,Nakatsuji:2000aa,Nakatsuji:2002aa,Yasuda:2002aa} In second quantization, the CSE is written in terms of its Hermitian and anti-Hermitian parts and vanishes at any eigenstate of $H$,
\begin{equation}
0 = \langle \psi|\{a^\dagger_ia^\dagger_ja_la_k,(H-E)\}|\psi\rangle + \langle \psi|[a^\dagger_ia^\dagger_ja_la_k,(H-E)]|\psi\rangle,
\end{equation}
for all $i,j,k,l$,  where $\{\cdot,\cdot\}$ and $[\cdot,\cdot]$ denote the anti-commutator and commutator, respectively. 
More recently, others have reported using only the anti-Hermitian part of the CSE (ACSE) to find approximate solutions to the SE.~\cite{Mazziotti:2004aa,Mazziotti:2007ad} While the CSE implies the SE, the ACSE does not necessarily imply the SE. Nonetheless, the ACSE has been shown to be highly effective in describing all-electron correlation in a wide variety of chemical contexts.~\cite{Boyn:2021aa,Foley:2021aa,Gidofalvi:2009aa,Snyder:2010aa,Greenman:2010aa,Snyder:2011aa,Foley:2009aa,Greenman:2011aa,Mazziotti:2008aa}

Practical solution of the ACSE requires minimizing a residual of the commutator,
\begin{equation}
    R^{ij}_{kl} = \langle \psi|[a^\dagger_ia^\dagger_ja_la_k,H]|\psi\rangle.
    \label{eq:residual_2quant}
\end{equation}
Using the canonical anti-commutation relations for electrons. The residual equation can be written in terms of the 2- and 3-RDMs,
\begin{equation}
\begin{aligned}
    R^{ij}_{kl} &= \sum_{pqr}  2K^{ij}_{qp}~{}^2D^{kl}_{qp}+ 6K^{kp}_{rq}~{}^3D^{ijp}_{rql}-6K^{lp}_{rq}~{}^3D^{ijp}_{rqk} \\ 
    &-2K^{kl}_{qp}~{}^2D^{ij}_{qp}  -6K^{pq}_{ri}~{}^3D^{jpq}_{rlk}+6K^{pq}_{rj}~{}^3D^{ipq}_{rlk}.
\end{aligned}
\label{eq:residual}
\end{equation}
where the elements of the 2-RDM ($^2D$) and 3-RDM ($^3D$) are defined by,
\begin{equation}
    \begin{aligned}
        ^2D^{ij}_{kl} &= \frac{1}{2}\langle \psi |a^\dagger_ia^\dagger_ja_la_k|\psi\rangle \\
        ^3D^{ijk}_{lmn} &= \frac{1}{6}\langle \psi | a^\dagger_i a^\dagger_j a^\dagger_k a_n a_m a_l |\psi \rangle.
    \end{aligned}
\end{equation}
The trace conventions for all relevant RDMs are detailed in the Appendix.

Expression of the ACSE in terms of the 3-RDM raises several issues. First, as the 3-RDM's memory requirements scale as $\mathcal{O}(r^6)$ with the number of orbitals, it is impractical to be stored in memory for all but the smallest systems. More fundamentally, the dependence on the 3-RDM results in an indeterminate system of equations.~\cite{Harriman:1979aa,Yasuda:2002aa,Alcoba:2005aa,DePrince:2007aa} In order to ameliorate both problems, the ACSE is written in terms of only the 2-RDM by employing approximate reconstructions of the 3-RDM and solved through an iterative procedure. Considering the full expression for the 3-RDM,
\begin{equation}
    ^3D = (3
    {^2\Delta} + ({^1D} \wedge  {^1D})) \wedge {^1D} + {^3\Delta},
\end{equation}
where $^1D$ is the 1-RDM, $\wedge$ indicates the antisymmetric wedge product, and $^2\Delta$ and $^3\Delta$ are the 2- and 3-body cumulants, respectively. The simplest approximation, proposed by Valdemoro, directly neglects the $^3\Delta$ term, resulting in the V reconstruction functional.~\cite{Colmenero:1993aa} At least two other reconstruction functionals are known, derived by Nakatsuji and Yasuda (NY), and Mazziotti (M).~\cite{Nakatuji:1996aa,Mazziotti:1999aa,Mazziotti:2000aa,DePrince:2007aa} The NY reconstruction is defined as:
\begin{equation}
    \begin{aligned}
        {}^3\Delta^{ijp}_{rql} = \frac{1}{6}\sum_{a}\sigma_{a}\hat{A}({}^2\Delta^{ia}_{rq}{}^2\Delta^{jp}_{al})\,,
    \end{aligned}
\end{equation}
where $\sigma_a$ is 1 if $a$ is an occupied orbital and -1 if it is an unoccupied orbital in the Hartree-Fock reference. $\hat{A}$ is the antisymmetry operator, which here permutes all indices except $a$. In this work, we examine only the V and NY functionals. We note that the NY reconstruction is derived with respect to a Hartree-Fock wavefunction, so its use with MR wavefunctions warrants detailed investigation.~\cite{DePrince:2007aa}

Even with reconstruction functionals, solving the ACSE requires extra care because direct minimization of the residual will result in an unphysical 2-RDM that is not $N$-representable. Instead, recognizing that the residual contains information about the gradient, we employ an iterative solution to update the 2-RDM via an Euler step,
\begin{equation}
\begin{aligned}
    {^2U}^{ij}_{kl} &= \langle \psi|[a^\dagger_ia^\dagger_ja_la_k,R]|\psi \rangle \\
    {^2D_{n+1}} &= {^2D}_n + \epsilon ~{^2U},
\end{aligned}
\label{eq:update}
\end{equation}
where ${^2D_n}$ is the 2-RDM at the $n^{th}$ iteration, and $\epsilon$ is a small parameter.~\cite{Mazziotti:2006aa} Other methods have been used to minimize the ACSE or CSE residuals, including a finite difference, quasi-Newton, direct inversion of the iterative subspace (DIIS), and purification approaches.~\cite{Sand:2015aa,Smart:2022,Colmenero:1994aa,Nooijen:2000aa,Alcoba:2005aa} A major challenge for ACSE, and RDM-based approaches in general, is maintaining a physical $N$-representable density matrix throughout the minimization.~\cite{Valdemoro:2000aa,Nooijen:2000aa}

Either a single- or multi-reference wavefunction can be used as a reference for the ACSE, but multi-reference states may have large elements of $^3\Delta$ when all indices correspond to active orbitals.~\cite{Mazziotti:2007aa} While clearly violating the V reconstruction, this can be mitigated by setting the elements of the residual to zero when their indices are all active. Here, we either include the elements of fully active indices of the residual, denoted as True, or set them to zero, denoted as False. The restriction of the active-active propagation of the ACSE is essential when using the V reconstruction; however, the behavior of this constraint has not been explored in the context of other reconstructions.


For a wavefunction with fixed spin, there are three unique spin blocks of the residual, namely $\alpha \alpha$, $\alpha \beta$, and $\beta \beta$. For singlets only two blocks are required because the $\alpha \alpha$ and $\beta \beta$ blocks are equivalent. In the Appendix, we provide explicit spin-free contractions for the residual using the V and NY reconstructions, and representative spin-block contractions. In each case, the bottleneck contraction is $\mathcal{O}(r^6)$ where $r$ is either the spin or spatial orbital dimension. Currently, these contractions are implemented using the numpy einsum function, which computes the optimal contraction path on the fly.~\cite{numpy} For a general non-singlet case, using the Valdemoro or NY reconstructions, we require 20 or 96 $\mathcal{O}(r^6)$ tensor contractions, respectively, to compute the residual norm.

\begin{figure}
    \centering
    \includegraphics[width=\linewidth]{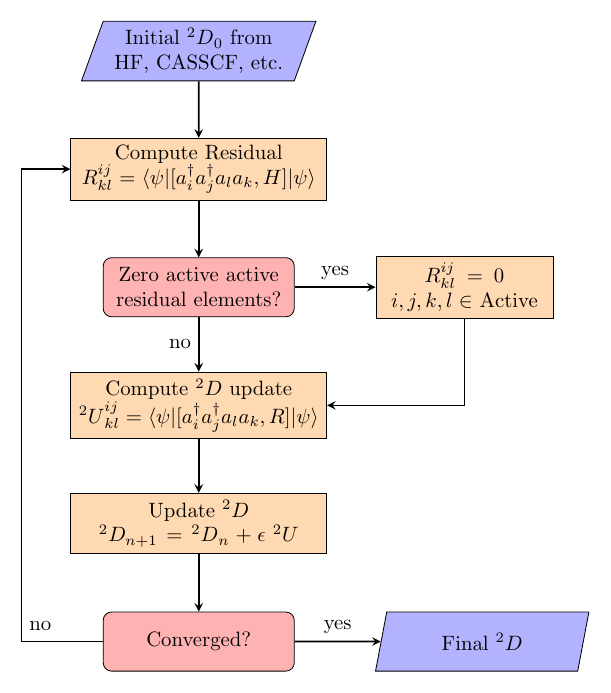}
    \caption{Diagram of the ACSE workflow.}
    \label{fig:placeholder}
\end{figure}

\section{Methods}

Our implementation of the ACSE is written in Python and publicly available on github.\cite{ACSE_github} It interfaces with PySCF for the CASSCF calculations to generate the initial RDMs and electron integrals.~\cite{Sun:2020aa} Currently, the V and NY reconstructions are available. Tensor contractions are implemented using einsum from the numpy library,~\cite{numpy} yielding $\mathcal{O}(r^6)$ scaling with respect to the number of orbitals. The working equations are implemented in the spatial-orbital basis, and we store the $\alpha\alpha$, $\alpha\beta$, and $\beta\beta$ 2-RDM blocks. Unless otherwise specified we choose $\epsilon = 1\times10^{-3}$ for the step size of the update in Eq.~\ref{eq:update}. In our algorithm, the ACSE iterations are terminated if there is an increase in the energy or the residual norm, or if the change in energy is less than $1\times10^{-6}$ H;~\cite{Mazziotti:2007aa} however, other stopping criteria have also been reported.~\cite{Mazziotti:2006aa,Mazziotti:1998aa}

We compare the ACSE results with SC-NEVPT2 as implemented in PySCF,~\cite{Angeli:2001aa,Angeli:2001ab,Angeli:2002aa} and benchmark both methods against exact FCI or DMRG-FCI. DMRG calculations were performed with block2 interfaced with PySCF.~\cite{Zhai:2021aa,Zhai:2023aa} The main parameter that determines the accuracy of a DMRG calculation is the bond dimension, which we denote as $M$, and we compute energy extrapolations with respect to $M$ where appropriate.~\cite{Olivares-Amaya:2015ab} We employ a variety of basis sets, namely, 6-31G,\cite{Hehre:1972aa,Hariharan:1973aa,Hariharan:1974aa,Ditchfield:1971aa} cc-pVDZ,\cite{Dunning:1989aa} cc-pVTZ,\cite{Kendall:1992aa} def2-SVP, and def2-TZVP.\cite{Weigend:2005aa,Weigend:2006aa}. All active space calculations are denoted $[n_e,n_o]$, for $n_e$ and $n_o$ active electrons and orbitals, respectively.

\section{Results and discussion}

We first examine the performance of the ACSE for the symmetric dissociation of linear H$_6$ in 6-31G, cc-pVDZ, and cc-pVTZ bases, using a [6,6] CASSCF reference. Table~\ref{tab:H6} shows the error in mH with respect to FCI using ACSE and NEVPT2 for three points along the dissociation: near equilibrium (0.9 \AA), in the bond-breaking region (1.4 \AA), and in the dissociated limit (5.0 \AA). As demonstrated from the mean signed error (MSE), both the V and NY reconstructions perform well compared to NEVPT2, with important exceptions. For example, in the bond-breaking region, the V reconstruction displays relatively large errors when active-active propagation is allowed (True). At 1.4 \AA~the highest occupied and lowest unoccupied natural orbital (HONO and LUNO) occupations are 1.73 and 0.27, respectively, indicating the importance of multireference correlation at that point. 
\begin{table}
    \centering
    \begin{tabular*}{9cm}{@{\extracolsep{\fill}}ccccccc} 
\hline \hline
& & \multicolumn{4}{c}{ACSE} & \\ \cline{3-6}
Basis & R(\AA) & V False & V True & NY False & NY True & NEVPT2 \\
\cline{1-1} \cline{2-2} \cline{3-3} \cline{4-4} \cline{5-5} \cline{6-6} \cline{7-7}
        &   0.9 & -0.62     & -4.38     & 0.84     & 0.55       & 5.31 \\
6-31G   &   1.4 & -5.07     & -18.64    & -2.19    & -3.41      & 1.71 \\
        &   5.0 & 0.00      & 0.00      & -9.14    & -1.14      & 0.00 \\
        &   MSE & -1.90     & -7.67     & -3.50    & -1.33      & 2.34 \\
        \hline
        &   0.9 & -1.58     & -4.49     & 1.54     & 0.90       & 17.04 \\
cc-pVDZ &   1.4 & -5.08     & -16.75    & -3.00    & -3.27      & 8.97 \\
        &   5.0 &  0.00     &  0.00     & -10.72   & -1.16      & 0.00 \\
        &   MSE & -2.22     & -7.08     & -4.06    & -1.18      & 8.67 \\
        \hline
        &   0.9 & -5.89     & -7.62     &  1.56    & 0.55       & 21.50 \\
cc-pVTZ &   1.4 & -5.79     & -15.58    &  -0.77   & -2.57      & 13.17 \\
        &   5.0 & 0.02      & 0.02      & -6.45    & -1.11      & 0.00 \\
        &   MSE & -3.89     & -7.73     & -1.19    & -1.04      & 11.56 \\
        \hline \hline
    \end{tabular*}
        \caption{ACSE and NEVPT2 error (mH) with respect to FCI for linear H$_6$ at three interatomic distances and using three basis sets.}
    \label{tab:H6}
\end{table}

As noted above, we expect the error in the V reconstruction to be amplified when active-active propagation is allowed. Furthermore, the NY reconstruction exhibits relatively large error at dissociation when active-active propagation is not allowed (False). On the other hand, on average the NY reconstruction provides the best results for all methods when active-active propagation is allowed (True). The results in Table~\ref{tab:H6} also reveal a subtlety when using the NY reconstruction and active-active propagation (True), namely, that in the intermediate region (1.4 \AA) the relative error increases compared to the equilibrium and dissociated limits. This is somewhat expected because the NY functional is explicitly defined with respect to a Hartree-Fock state, so the behavior of the reconstruction for multireference states may be unreliable. We note that the average error of NEVPT2 increases with the basis size, while the error with ACSE is relatively stable. This demonstrates that the ACSE generally recovers more correlation energy than NEVPT2, but it is important to note that NEVPT2 still yields accurate relative energies for this system.

\begin{figure}[h]
    \centering
    \includegraphics[width=\linewidth]{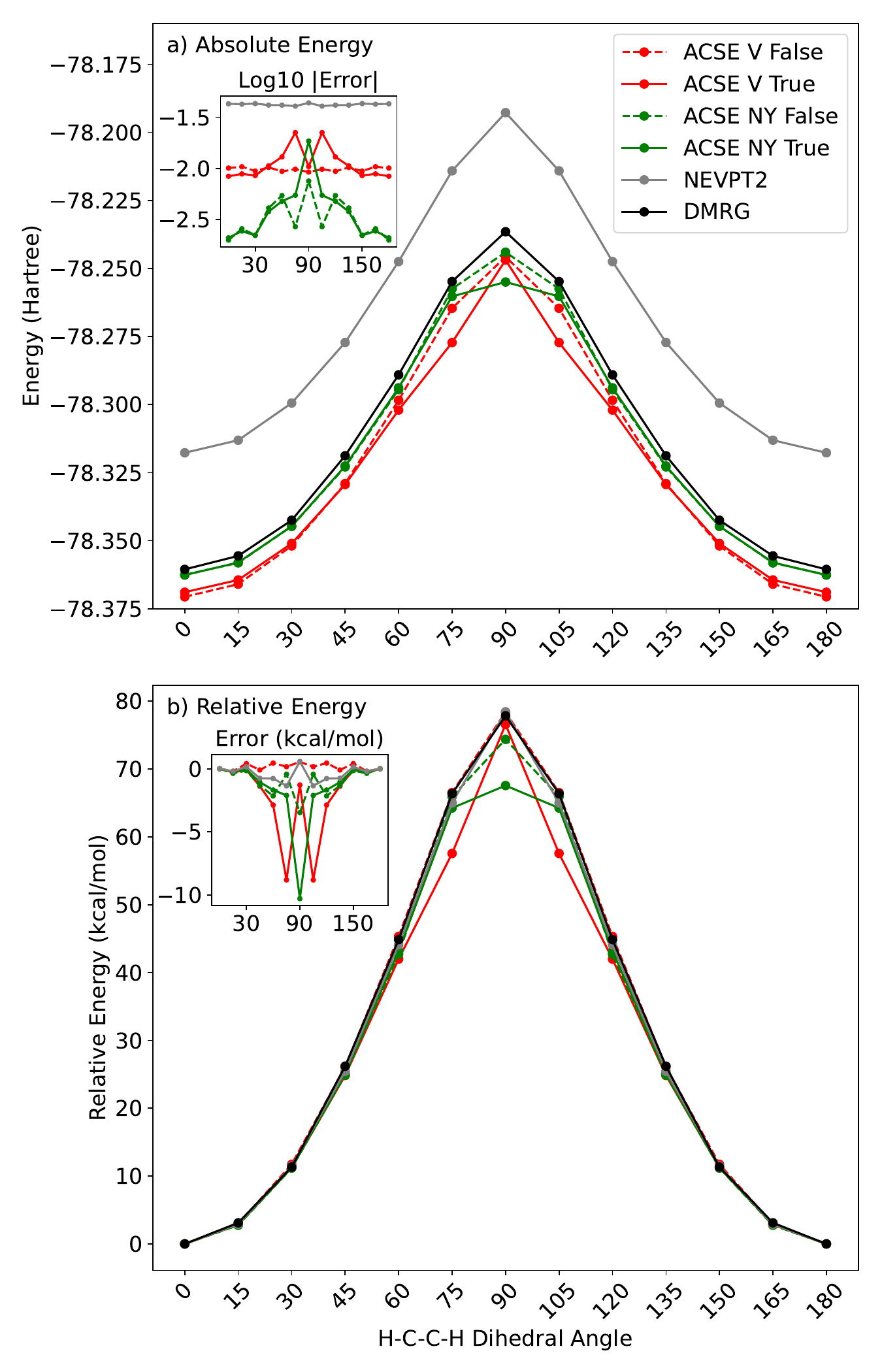}
    \caption{Ethylene (C$_2$H$_4$) dihedral barrier height a) absolute energies (H) and b) the relative energies (kcal/mol) using extrapolated DMRG-FCI ($M=1100,1300,1500$), NEVPT2, and ACSE in a cc-pVDZ basis. The top inset shows the log of the absolute error, and the bottom inset shows the relative error.}
    \label{fig:ethylene_combined}
\end{figure}

The comparison of absolute and relatively error for these methods is further elucidated by examining the barrier for C$_2$H$_4$ bond rotation with respect to the H-C-C-H dihedral angle. Figure~\ref{fig:ethylene_combined}a and Figure~\ref{fig:ethylene_combined}b show the absolute and relative energies, respectively, of the ACSE variants and NEVPT2 compared to DMRG-FCI in the cc-pVDZ basis with an [8,8] active space. Here, we extrapolate the DMRG results using the bond dimensions $M=1100,1300,1500$ to approximate the exact FCI result.~\cite{Olivares-Amaya:2015ab} Fig.~\ref{fig:ethylene_combined}a clearly shows a consistent and significant absolute error for NEVPT2 at about 40 mH above the DMRG-FCI energy. In contrast, the ACSE results are consistently about an order of magnitude more accurate than NEVPT2 for absolute energies, as seen in the inset of Fig.~\ref{fig:ethylene_combined}a.

The relative errors shown in Fig.~\ref{fig:ethylene_combined} demonstrate the accuracy of both ACSE variants and NEVPT2 throughout the rotation. Indeed, both NEVPT2 and the ACSE with the V reconstruction and no active-active propagation (False) yield highly accurate potential energy surfaces. Notably, the V (False) result has a maximum error of only 0.54 kcal/mol, compared to NEVPT2 with a maximum error of about -1.35 kcal/mol. Importantly, the error for V (False) is stable throughout the dihedral rotation, which demonstrates robust behavior through transitions from weakly correlated to strongly correlated regimes. However, here, we again see unreliable behavior from the NY reconstruction away from weakly correlated states. As the rotation angle increases and the state becomes more multireference, the relative error of the NY reconstruction generally increases as well. For both the NY and V reconstructions, the inclusion of active-active propagation (True) is detrimental to the accuracy of the ACSE in this case, most significantly for NY (True), which significantly underestimates the barrier height by about 10 kcal/mol.

\begin{figure}
    \centering
    \includegraphics[width=\linewidth]{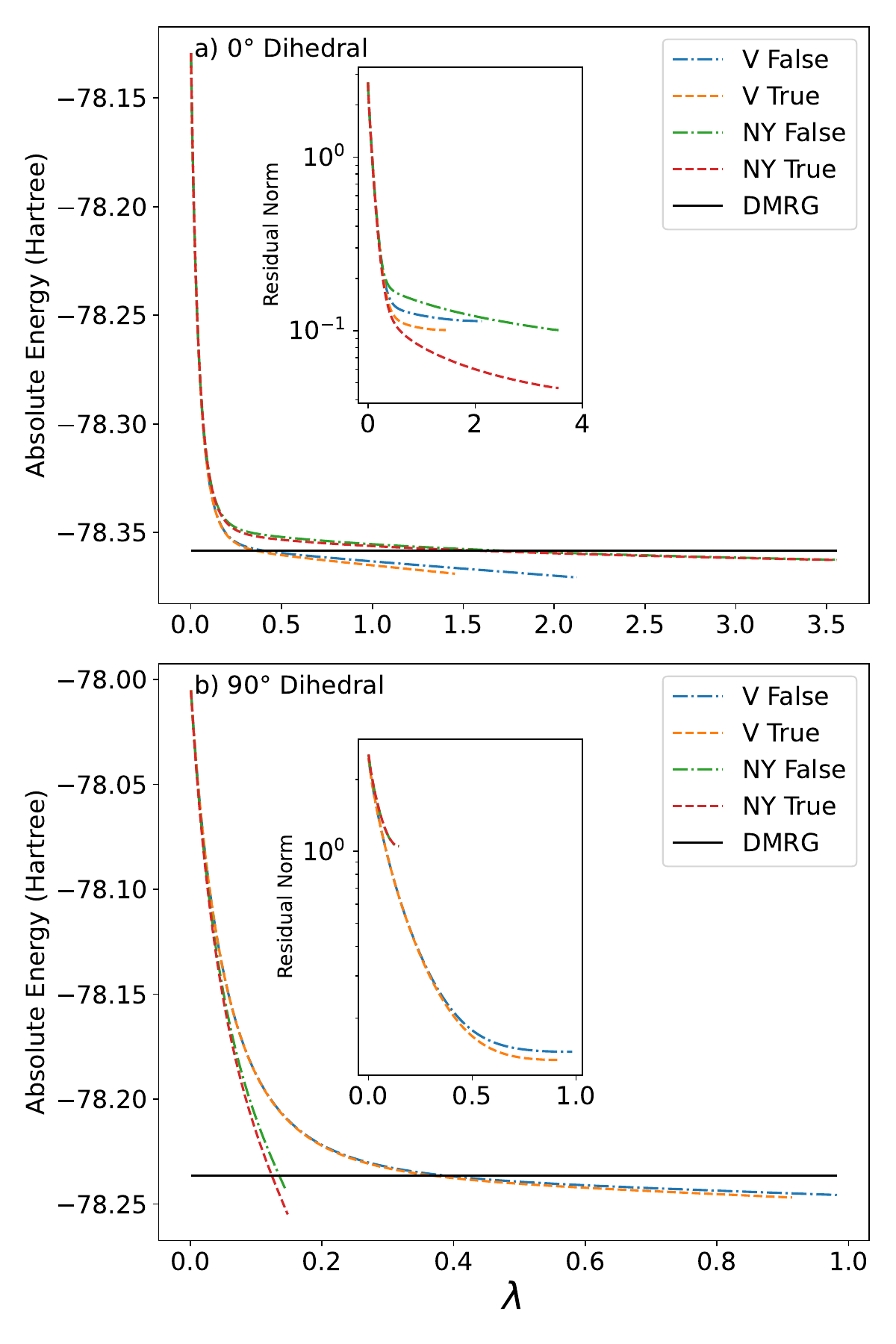}
    \caption{Energy convergence (in Hartree) and log$_{10}$ residual norm convergence vs $\lambda=\epsilon n$ for the 0 and 90\degree~dihedral angles of ethylene.}
    \label{fig:ethylene_0}
\end{figure}

Due to the approximate update, it is important to monitor the convergence of both the energy and the residual norm in ACSE calculations. To survey different regimes of correlation, we analyze the convergence of the ethylene rotation when the dihedral is 0\degree ~and 90\degree, where the former is weakly correlated and the latter is more strongly correlated. Here, we define $\lambda=\epsilon n$, where $\epsilon$ is the step size from Eq. \ref{eq:update} and $n$ is the iteration number. Figure \ref{fig:ethylene_0}a shows the convergence of the energy over the trajectory, along with the convergence of the residual norm in the inset, for ethylene with a 0\degree ~dihedral angle. The shape of the trajectory is in general an exponentially damped region, followed by an asymptotic regime. After the damped region, the ACSE methods split into two pairs of curves depending on the reconstruction. In this case, the lower energy curves are the result of the V reconstruction, while the inset shows the V reconstructions are essentially stationary in the residual norm near $\lambda=2$. The two higher energy curves denote the NY reconstruction, which provides better agreement with the DMRG-FCI energy. Beyond $\lambda$=0.5 the change in energies for both reconstructions is approximately linear with slopes of -8 and -5 mH/$\lambda$ for V and NY, respectively. Comparison of the residual norm indicates that the NY reconstruction is achieving a stable, and likely better solution than V, in light of the smaller norm. There is a clear difference in the norms if active-active propagation is allowed,  where the (False) calculations always achieve a larger norm because of the restriction of degrees of freedom. We note that the residual norms are calculated before setting those elements of the residual to zero to enable comparison.

At the strongly correlated 90\degree~ dihedral angle, the convergence behavior of the ACSE reveals significant differences depending on the details of the algorithm, shown in Figure \ref{fig:ethylene_0}b. Again, characterized by an initial exponential damping, the V results proceed to a linear regime, while the NY results diverge. The energy convergence for V is stable, but it again overestimates the correlation energy. In contrast, the NY reconstruction never reaches the asymptotic regime of the energy trajectory, and instead the calculation is terminated due to the increase in the residual norm, shown in the inset. Clearly, the NY calculations are not converged to an eigenstate, which provides a rationale for the large error shown in Fig.~\ref{fig:ethylene_combined} for NY at 90\degree. We hypothesize that this is due to the assumption of a Hartree-Fock style cumulant reconstruction, which could result in unstable convergence. Analysis of the convergence of ACSE calculations provides vital information about the success of the algorithm in approximating solutions to the Schr\"odinger equation.


To examine the performance of the ACSE for excited states we compute the lowest energy singlet transition of ethylene, S$_0$ to S$_1$, compare to NEVPT2, and benchmark against DMRG-FCI. Table~\ref{tab:ethylene_excited} shows the results for the 6-31G, cc-pVDZ, and cc-pVTZ basis sets. For the 6-31g and cc-pVDZ DMRG-FCI calculations, we select $M=1000$, and for cc-pVTZ, we select $M=800$. Following previous work, we perform equally-weighted state averaged CASSCF calculations for [2,2] and [6,6] active spaces.~\cite{Neese2025} The [2,2] active space contains only the $\pi$ and $\pi^*$ orbitals, while the [6,6] active space includes weakly correlated $\sigma$-type orbitals, allowing us to examine how the active space composition alters the performance of the ACSE. As expected, the inclusion of dynamical correlation via NEVPT2 or ACSE significantly reduces the errors compared to the CASSCF result. The V reconstruction produces excellent results using the smaller active space reference, yielding average errors less than 10 meV compared to DMRG-FCI, and a significant improvement over NEVPT2 with average error of 250 meV. The NEVPT2 results are excellent using the larger active space, while the ACSE results worsen considerably. Unfortunately, NY yields poor results for both active spaces and for both orbital-propagation protocols. This increase in error is in line with our previous observations -- the NY reconstruction fails in the strongly correlated regime, where occupations deviate significantly from the Hartree-Fock reference of either 0 or 2, which is the case in the excited state with occupations of nearly 1 for both the HONO and LUNO. For the V reconstruction, the ACSE with the larger active space introduces significant errors, even when orbital propagation is restricted.

\begin{table*}[t]
    \centering
    \begin{tabular*}{\textwidth}{@{\extracolsep{\fill}}ccccccc cccccc c} 
    \hline \hline
        & \multicolumn{6}{c}{[2,2] Active Space} & \multicolumn{6}{c}{[6,6] Active Space} & FCI\\ 
        \cline{2-7} \cline{8-13} \cline{14-14} 
        Basis & V False & V True & NY False & NY True & CASSCF & NEVPT2 & V False & V True & NY False & NY True & CASSCF & NEVPT2 & DMRG \\
        \cline{1-1} \cline{2-2} \cline{3-3} \cline{4-4} \cline{5-5} \cline{6-6} \cline{7-7} \cline{8-8} \cline{9-9} \cline{10-10} \cline{11-11} \cline{12-12} \cline{13-13} \cline{14-14}
	6-31G	&	-0.04	&	-0.04	&	-0.11	&	-0.86	&	1.18	&	-0.21	&	0.42	&	0.78	&	0.22	&	0.20	&	0.48	&	0.08	&	9.38$^a$	\\
	cc-pVDZ	&	-0.17	&	-0.17	&	-0.17	&	-0.72	&	1.03	&	-0.37	&	0.26	&	0.60	&	-0.08	&	-0.47	&	0.51	&	-0.11	&	8.89$^a$	\\
	cc-pVTZ	&	-0.04	&	-0.05	&	0.41	&	0.02	&	1.01	&	-0.18	&	0.45	&	0.73	&	0.39	&	0.32	&	0.66	&	0.01	&	8.36$^b$	\\
	MUE	&	0.08	&	0.09	&	0.23	&	0.53	&	1.07	&	0.25	&	0.38	&	0.70	&	0.23	&	0.33	&	0.55	&	0.07	&		- \\
        
    \hline \hline
    \end{tabular*}
    \caption{Error in the ethylene S$_0$ to S$_1$ excitation energy (eV) with respect to DMRG-FCI calculations for CASSCF, NEVPT2, and ACSE. ${}^a$ $M=1000$, ${}^b$ $M=800$.}
    \label{tab:ethylene_excited}
\end{table*}


These results highlight the effects of allowing active-active propagation of the ACSE, along with subtleties in the choice of reconstruction. For the V reconstruction, nearly all of the results are worse when the active-active elements contribute to the propagation. As expected, the error is worse in strongly correlated domains, where large elements of $^3\Delta$ are ignored by the V reconstruction. For the NY reconstruction, the situation is more complicated because NY approximates the $^3\Delta$, but is defined with respect to a Hartree-Fock reference state. For NY, the active-active propagation (True) is generally more accurate, except in strongly correlated regimes, where NY may not be an appropriate reconstruction for the multi-configurational reference. A reconstruction that approximates $^3\Delta$ and is defined with respect to a multi-configurational reference, such as the reconstruction of Mazziotti, will likely exhibit the most robust performance in correlated regimes, without restricting the active-active degrees of freedom.~\cite{Mazziotti:2007aa,DePrince:2007aa}


\begin{figure}[h]
    \centering
    \includegraphics[width=\linewidth]{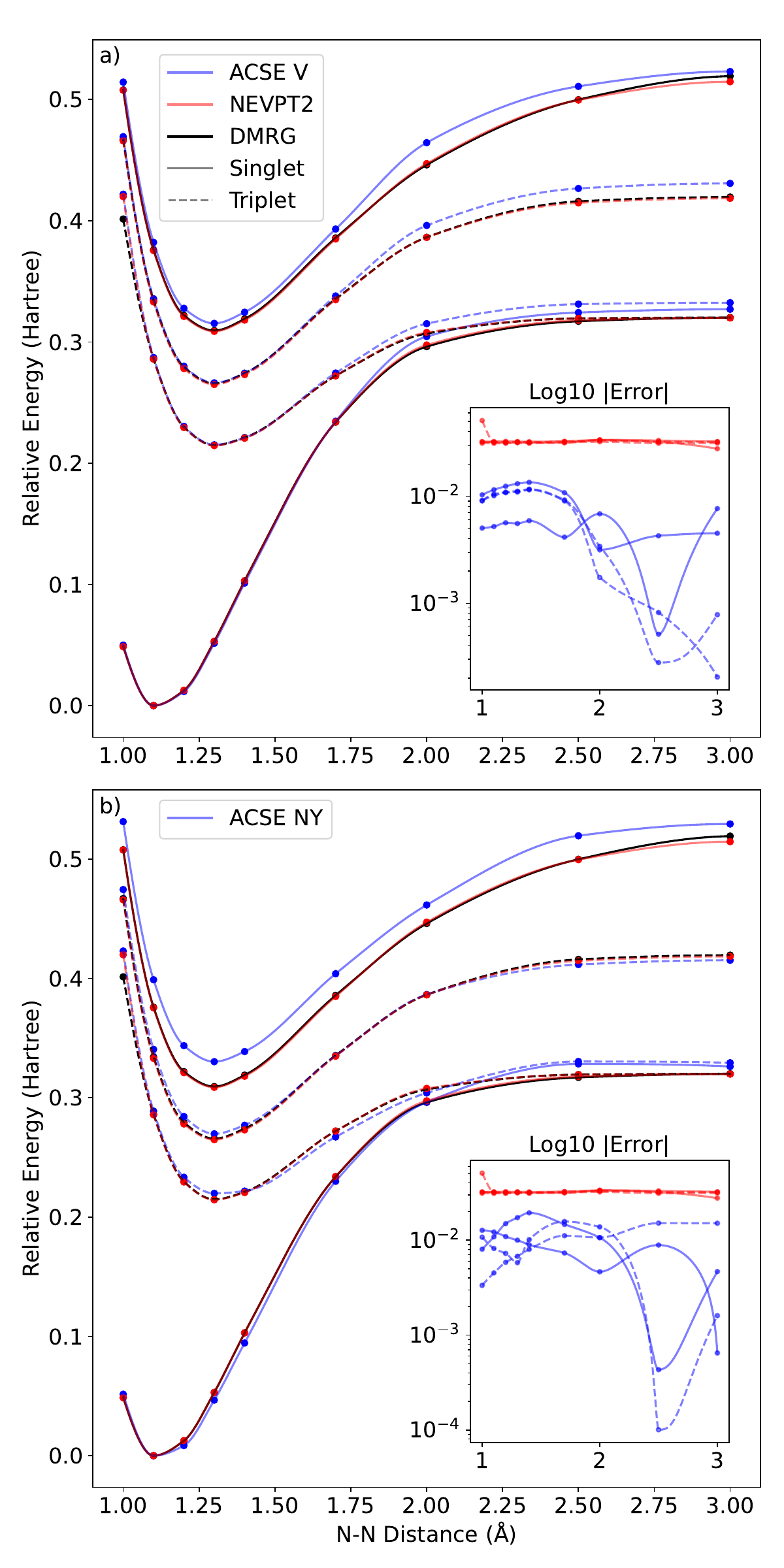}
    \caption{The dissociation of the two lowest energy singlets and triplets of N$_2$ in a cc-pVDZ basis set, computed with NEVPT2 and ACSE with a [6,6] CASSCF reference, and DMRG-FCI ($M=1200$). The insets show the log of the absolute error with respect to DMRG-FCI}
    \label{fig:N2_combined}
\end{figure}

Additionally, we explore the ability of the ACSE to describe different spin states in both single-reference and multi-reference regimes. We first calculate the two lowest energy singlet and triplet states along the dissociation of molecular nitrogen, presented in Figure \ref{fig:N2_combined}. For the N$_2$ calculations, we use the V and NY reconstructions without active-active propagation (False) and employ a [6,6] CASSCF active space for the ACSE and NEVPT2 calculations. For this system, NEVPT2 yields good agreement along the potential energy surface, with average relative errors for the S$_0$, S$_1$, T$_0$, and T$_1$ states of 0.18, 0.69, 1.47, and 0.58 kcal/mol, respectively. The result of the ACSE with the V reconstruction is shown in Figure \ref{fig:N2_combined}a, which provides excellent reproduction of the S$_0$, T$_0$, and T$_1$ states around their respective minima with maximum relative errors of 0.74, 0.36, and 0.44 kcal/mol, respectively. As the system becomes more multi-configurational with increasing bond length, these errors increase to 5-7 kcal/mol. Comparison of the absolute errors vs the DMRG-FCI reference, presented in the inset of Figure \ref{fig:ethylene_combined}a, confirms that the absolute errors obtained with the ACSE are always lower than those obtained from NEVPT2, while both methods provide qualitatively very similar dissociation curves.

Employing the NY reconstruction, presented in Figure \ref{fig:N2_combined}b, we see that the accuracy is in general worse compared to the V reconstruction. The maximal relative errors around the S$_0$, T$_0$, and T$_1$ minima are larger, around 3 kcal/mol, and again the relative error increases with increasing bond length for the S$_0$ and T$_0$ states to a maximum of about 7 kcal/mol, respectively. Conversely, the errors for the T$_1$ state are smaller with a maximal error throughout dissociation of about 3 kcal/mol. Finally, the S$_1$ state reveals consistent and larger relative errors with the NY reconstruction compared to the V reconstruction with average relative errors of approximately 12 kcal/mol and a maximal error of 15 kcal/mol. Comparison of the absolute errors vs the DMRG-FCI reference with the NY reconstruction, presented in the inset of Figure \ref{fig:ethylene_combined}b, again confirms that the ACSE MAEs are lower than those obtained from NEVPT2. The MAEs for the ACSE for the NY reconstruction are 7.04, 5.32, 5.12, and 5.62 kcal/mol for the S$_0$, S$_1$, T$_0$, and T$_1$ states, respectively, compared to 1.22, 2.08, 0.46, and 1.14 kcal/mol, respectively, compared to the V reconstruction. While the NY reconstruction has less impressive results than the V reconstruction, it still recovers qualitatively correct dissociation behavior in all areas of the potential surface. Nonetheless, these results confirm that the NY reconstruction is less robust than the V reconstruction when considering strongly correlated systems, as well as with restricted active-active propagation.

Finally, we use the ACSE to compute atomic spin state energies for the singlet, triplet, and quintet states of Fe$^{2+}$ and Co$^{3+}$, along with the doublet, quartet, and sextet states of Fe$^{3+}$ in the def2-SVP and def2-TZVP basis sets. We employ a minimal active space comprised of the 3$d$ orbitals, resulting in a [6,5] active space for Fe$^{2+}$ and Co$^{3+}$, and a [5,5] active space for Fe$^{3+}$.~\cite{Zhang:2020} Our calculations omit spin-orbit coupling, so we present the experimental results with spin-orbit coupling effects removed via the first-order method.\cite{Zhang:2020,Phillips:2011aa,NIST} For the ACSE calculations we found $\epsilon=1\times 10^{-5}$ was necessary for stable propagation. For the atomic calculations we exclusively use the V reconstruction without active-active propagation (False), and compare with CASSCF and NEVPT2.

\begin{table}[]
    \centering
    \begin{tabular*}{9cm}{@{\extracolsep{\fill}}ccccccc} 
    \hline \hline
   Atom & Method & Basis &3$\xrightarrow{}$1 & 5$\xrightarrow{}$1 & 5$\xrightarrow{}$3 & MUE    \\
   \cline{1-1} \cline{2-2} \cline{3-3} \cline{4-4} \cline{5-5} \cline{6-6} \cline{7-7}
\multirow{8}{*}{Fe$^{2+}$}	&	 \multirow{3}{*}{CASSCF} 	&	 def2-SVP             	&	2.21	&	11.58	&	9.36	&	7.72	\\
                          	&	                         	&	 def2-TZVP            	&	1.33	&	10.81	&	9.47	&	7.20	\\
                          	&	 \multirow{2}{*}{NEVPT2} 	&	 def2-SVP             	&	2.42	&	7.97	&	5.54	&	5.31	\\
                          	&	                         	&	 def2-TZVP            	&	-2.08	&	2.79	&	4.86	&	3.24	\\
                          	&	 \multirow{2}{*}{ACSE}   	&	 def2-SVP             	&	0.37	&	5.37	&	4.99	&	3.58	\\
                          	&	                         	&	 def2-TZVP            	&	1.47	&	2.14	&	0.63	&	1.41	\\
                          	&	 Experiment              	&	   -                   	&	29.52	&	85.58	&	56.07	&	-     \\\hline
\multirow{8}{*}{Co$^{3+}$}	&	 \multirow{3}{*}{CASSCF} 	&	 def2-SVP             	&	1.44	&	10.91	&	9.48	&	7.28	\\
                          	&	                         	&	 def2-TZVP            	&	1.17	&	10.20	&	9.02	&	6.80	\\
                          	&	 \multirow{2}{*}{NEVPT2} 	&	 def2-SVP             	&	-0.39	&	5.65	&	6.04	&	4.03	\\
                          	&	                         	&	 def2-TZVP            	&	-1.00	&	4.04	&	5.04	&	3.36	\\
                          	&	 \multirow{2}{*}{ACSE}   	&	 def2-SVP             	&	0.14	&	6.60	&	6.46	&	4.40	\\
                          	&	                         	&	 def2-TZVP            	&	-1.16	&	3.67	&	4.82	&	3.22	\\
                          	&	 Experiment              	&	   -                  	&	34.76	&	101.00	&	66.24	&	-	\\\hline
      	&	  	&	  	&	4$\xrightarrow{}$2 	&	 6$\xrightarrow{}$2 	&	 6$\xrightarrow{}$4 & MSE \\
        \cline{4-4} \cline{5-5} \cline{6-6}  	\cline{7-7}
\multirow{8}{*}{Fe$^{3+}$}	&	 \multirow{3}{*}{CASSCF} 	&	 def2-SVP             	&	4.17	&	19.70  &	15.52	&	13.13	\\
                          	&	                         	&	 def2-TZVP            	&	3.80	&	18.53	&	14.72	&	12.35	\\
                          	&	 \multirow{2}{*}{NEVPT2} 	&	 def2-SVP             	&	3.64	&	12.61	&	8.96	&	8.40	\\
                          	&	                         	&	 def2-TZVP            	&	2.97	&	9.81	&	6.83	&	6.54	\\
                          	&	 \multirow{2}{*}{ACSE}   	&	 def2-SVP             	&	1.96	&	10.43	&	8.46	&	6.95	\\
                          	&	                         	&	 def2-TZVP            	&	-0.56	&	5.86	&	6.41	&	4.28	\\
                          	&	 Experiment              	&	   -                  	&	42.33	&	134.62	&	92.30	&	-        \\ \hline \hline
    \end{tabular*}
    \caption{Atomic spin flip energy errors with respect to experiment in kcal/mol. A [6,5] active space was used for Fe$^{2+}$ and Co$^{3+}$ and a [5,5] active space was used for Fe$^{3+}$. All ACSE calculations used the V reconstruction while restricting the propagation of the active-active orbitals (False).}
    \label{tab:atoms}
\end{table}

Table~\ref{tab:atoms} shows the error in computed atomic spin splittings compared to experiment in kcal/mol. For Fe$^{2+}$ CASSCF in the def2-SVP basis results in errors of 2.21, 11.58, and 9.36 kcal/mol for the triplet-singlet, quintet-singlet, and quintet-triplet spin splittings, respectively. Using the larger def2-TZVP basis results in minor improvements of about 1 kcal/mol. The ACSE in the def2-SVP basis notably improves upon CASSCF, resulting in errors of 0.37, 5.37, and 4.99 kcal/mol for the triplet-singlet, quintet-singlet, and quintet-triplet spin splittings, respectively. Increasing the basis set size to def2-TZVP results in an overall further reduction of these errors to 1.47, 2.14, and 0.63 kcal/mol, respectively, for excellent agreement with experiment. 


For Co$^{3+}$ (Table \ref{tab:atoms}), CASSCF again yields large errors compared to experiment, as expected. Using the larger basis sets again results in only minor changes in the computed spin splitting energies. The ACSE demonstrates high-quality results in the def2-SVP basis, reducing these errors to 0.14, 6.60, and 6.46 kcal/mol, respectively, while the larger def2-TZVP basis results in errors of -1.16, 3.67, and 4.82 kcal/mol. Again, these compare favorably with NEVPT2 in the larger basis, though NEVPT2 displays slightly lower errors for the def2-SVP basis for these transitions.

The results for Fe$^{3+}$ follow similar trends as above. CASSCF yields large errors for all states, which improve only slightly with the larger basis set. Employing the ACSE in the def2-SVP basis reduces these errors to 1.96, 10.43, and 8.46 kcal/mol for the quartet-doublet, sextet-doublet, and sextet-quartet, respectively. Here, the ACSE with def2-TZVP significantly reduces the error in the excitation energies with errors of -0.56, 5.86, and 6.41 kcal/mol, respectively. For this system NEVPT2 yields notably large errors even when using the larger basis set. These results further highlight the ability of the ACSE to resolve different spin multiplicities and yielding accurate spin splitting energetics.\cite{Gidofalvi:2009aa,Greenman:2011aa,Foley:2011aa,Schlimgen:2017aa,Boyn:2021aa}

\section{Conclusions}

We have developed an open source, python-based software for solving the ACSE, providing an interface to PySCF, enabling modularity and extendability.\cite{ACSE_github} Our initial implementation propagates the ACSE via an Euler step, efficiently calculates the residual via extensive use of numpy einsum functions, avoids explicit storage of the 3-RDM, and solves the ACSE effectively using only the 1- and 2-RDMs. Two 3-RDM reconstructions, Valdemoro (V) and Nakatsuji-Yasuda (NY), are implemented.\cite{Colmenero:1993aa,Nakatuji:1996aa} Employing this framework, we systematically benchmark several approximations to the ACSE employing a range of illustrative electronic systems, surveying ground and excited states in both dynamically correlated and strongly correlated regimes for main group and transition metal chemistry. The selection of benchmarks demonstrates that the ACSE can provide robust and accurate estimates of electronic states, and is competitive with the widely-adopted NEVPT2. Unlike NEVPT2, the complexity of the ACSE does not depend on the size of the active space, providing a potential advantage if the ACSE is coupled with large approximate active-space solvers.

As expected and reported previously, the convergence of the ACSE is sensitive to the specifics of the reconstruction and density-matrix propagation.~\cite{DePrince:2007aa} Generally, we find that while the NY reconstruction can yield good results in weakly correlated systems, it is likely not reliable when coupled with multireference initial wavefunctions. This remains true even when active-active propagation in the residual is prevented, which has previously been noted to improve ACSE performance in strongly correlated systems.\cite{Mazziotti:2007aa} The V reconstruction without active-active propagation provides excellent relative energies, often in line with or better than those obtained from NEVPT2 without the computational dependence on the active space 4-RDM, across a variety of chemical problems. Monitoring the convergence trajectory of the ACSE also provides valuable information about the about the stability of the putative solution. Future work will employ extrapolation and fitting techniques to estimate solutions to the ACSE if it stably enters the asymptotic regime of the trajectory.


Our open-source implementation of the ACSE, interfaced with PySCF, provides a functional code for testing and extending applications of contracted eigenvalue equations for quantum chemistry. The code requires few user-set algorithm parameters, and is directly interfaced with PySCF's CASSCF framework for direct comparison with other techniques available in the ecosystem. The ACSE provides an efficient way to find approximate 2-RDMs for strongly correlated electronic systems, and provides accurate and robust electronic energies. The algorithm complexity does not depend on the complexity of the reference wavefunction, nor does the ACSE require an effective Hamiltonian, such as the Dyall Hamiltonian used in NEVPT2. Further improvements to the ACSE include standard accelerators, such as the use of symmetry and parallelization, along with developmental advances, such as improved optimization and extrapolation techniques. In the context of MRPT and MRCC, we expect the python-based ACSE to provide a competitive and complementary approach to simulating many-electron correlation in ground and excited states.



\section{Data Availability}
The data that supports the findings of this study are available within the article. 

\begin{acknowledgments}

 The authors thank the University of Minnesota for start up funding, and the the Minnesota Supercomputing Institute (MSI) for computational resources.

\end{acknowledgments}

\bibliography{main} 

@article{Andersson:1990aa,
    Author = {Andersson, Kerstin. and Malmqvist, Per Aake. and Roos, Bjoern O. and Sadlej, Andrzej J. and Wolinski, Krzysztof.},
    Title = {Second-order perturbation theory with a {CASSCF} reference function},
    Journal = {The Journal of Physical Chemistry},
    Year = {1990},
    Volume = {94},
    Number = {14},
    Pages = {5483--5488},
    Month = {07},
}

@article{Andersson:1992aa,
    Author = {Andersson,Kerstin and Malmqvist,Per‐Åke and Roos,Björn O.},
    Title = {Second‐order perturbation theory with a complete active space self‐consistent field reference function},
    Journal = {The Journal of Chemical Physics},
    Year = {1992},
    Volume = {96},
    Number = {2},
    Pages = {1218--1226},
    Month = {2023/02/07},
}

@article{Angeli:2002aa,
    Author = {Angeli,Celestino and Cimiraglia,Renzo and Malrieu,Jean-Paul},
    Title = {{N}-electron valence state perturbation theory: A spinless formulation and an efficient implementation of the strongly contracted and of the partially contracted variants},
    Journal = {The Journal of Chemical Physics},
    Year = {2002},
    Volume = {117},
    Number = {20},
    Pages = {9138-9153},
}

@article{Guo:2016ab,
    Author = {Guo, Sheng and Watson, Mark A. and Hu, Weifeng and Sun, Qiming and Chan, Garnet Kin-Lic},
    Title = {N-Electron Valence State Perturbation Theory Based on a Density Matrix Renormalization Group Reference Function, with Applications to the Chromium Dimer and a Trimer Model of Poly(p-Phenylenevinylene)},
    Journal = {Journal of Chemical Theory and Computation},
    Year = {2016},
    Volume = {12},
    Number = {4},
    Pages = {1583--1591},
    Month = {04},
}

@article{Neese2025,
author = {Kempfer, Emily M. and Sivalingam, Kantharuban and Neese, Frank},
title = {Efficient Implementation of Approximate Fourth Order N-Electron Valence State Perturbation Theory},
journal = {Journal of Chemical Theory and Computation},
volume = {21},
number = {8},
pages = {3953-3967},
year = {2025},
}

@article{Nakatuji:1996aa,
  title = {Direct Determination of the Quantum-Mechanical Density Matrix Using the Density Equation},
  author = {Nakatsuji, Hiroshi and Yasuda, Koji},
  journal = {Phys. Rev. Lett.},
  volume = {76},
  issue = {7},
  pages = {1039--1042},
  numpages = {0},
  year = {1996},
  month = {Feb},
  publisher = {American Physical Society},
  doi = {10.1103/PhysRevLett.76.1039},
}

@article{Mazziotti:2000aa,
title = {Complete reconstruction of reduced density matrices},
journal = {Chemical Physics Letters},
volume = {326},
number = {3},
pages = {212-218},
year = {2000},
issn = {0009-2614},
doi = {https://doi.org/10.1016/S0009-2614(00)00773-9},
author = {David A. Mazziotti},
}

@article{Mazziotti:2007aa,
  title = {Multireference many-electron correlation energies from two-electron reduced density matrices computed by solving the anti-Hermitian contracted Schr\"odinger equation},
  author = {Mazziotti, David A.},
  journal = {Phys. Rev. A},
  volume = {76},
  issue = {5},
  pages = {052502},
  numpages = {8},
  year = {2007},
  month = {Nov},
  publisher = {American Physical Society},
  doi = {10.1103/PhysRevA.76.052502},
}

@Article{         numpy,
 title         = {Array programming with {NumPy}},
 author        = {Charles R. Harris and K. Jarrod Millman and St{\'{e}}fan J.
                 van der Walt and Ralf Gommers and Pauli Virtanen and David
                 Cournapeau and Eric Wieser and Julian Taylor and Sebastian
                 Berg and Nathaniel J. Smith and Robert Kern and Matti Picus
                 and Stephan Hoyer and Marten H. van Kerkwijk and Matthew
                 Brett and Allan Haldane and Jaime Fern{\'{a}}ndez del
                 R{\'{i}}o and Mark Wiebe and Pearu Peterson and Pierre
                 G{\'{e}}rard-Marchant and Kevin Sheppard and Tyler Reddy and
                 Warren Weckesser and Hameer Abbasi and Christoph Gohlke and
                 Travis E. Oliphant},
 year          = {2020},
 month         = sep,
 journal       = {Nature},
 volume        = {585},
 number        = {7825},
 pages         = {357--362},
 doi           = {10.1038/s41586-020-2649-2},
 publisher     = {Springer Science and Business Media {LLC}},
}

@article{Roelfes:2010,
  title={Artificial metalloenzymes},
  author={Rosati, Fiora and Roelfes, Gerard},
  journal={ChemCatChem},
  volume={2},
  number={8},
  pages={916--927},
  year={2010},
  publisher={Wiley Online Library}
}

@article{Ward:2018,
  title={Artificial metalloenzymes: reaction scope and optimization strategies},
  author={Schwizer, Fabian and Okamoto, Yasunori and Heinisch, Tillmann and Gu, Yifan and Pellizzoni, Michela M and Lebrun, Vincent and Reuter, Raphael and Kohler, Valentin and Lewis, Jared C and Ward, Thomas R},
  journal={Chemical reviews},
  volume={118},
  number={1},
  pages={142--231},
  year={2018},
  publisher={ACS Publications}
}

@article{Ward:2019,
  title={Artificial metalloenzymes: challenges and opportunities},
  author={Davis, Holly J and Ward, Thomas R},
  journal={ACS central science},
  volume={5},
  number={7},
  pages={1120--1136},
  year={2019},
  publisher={ACS Publications}
}

@article{Sun:2020aa,
    Author = {Sun, Qiming and Zhang, Xing and Banerjee, Samragni and Bao, Peng and Barbry, Marc and Blunt, Nick S. and Bogdanov, Nikolay A. and Booth, George H. and Chen, Jia and Cui, Zhi-Hao and Eriksen, Janus J. and Gao, Yang and Guo, Sheng and Hermann, Jan and Hermes, Matthew R. and Koh, Kevin and Koval, Peter and Lehtola, Susi and Li, Zhendong and Liu, Junzi and Mardirossian, Narbe and McClain, James D. and Motta, Mario and Mussard, Bastien and Pham, Hung Q. and Pulkin, Artem and Purwanto, Wirawan and Robinson, Paul J. and Ronca, Enrico and Sayfutyarova, Elvira R. and Scheurer, Maximilian and Schurkus, Henry F. and Smith, James E. T. and Sun, Chong and Sun, Shi-Ning and Upadhyay, Shiv and Wagner, Lucas K. and Wang, Xiao and White, Alec and Whitfield, James Daniel and Williamson, Mark J. and Wouters, Sebastian and Yang, Jun and Yu, Jason M. and Zhu, Tianyu and Berkelbach, Timothy C. and Sharma, Sandeep and Sokolov, Alexander Yu. and Chan, Garnet Kin-Lic},
    Title = {Recent developments in the PySCF program package},
    Journal = {The Journal of Chemical Physics},
    Year = {2020},
    Volume = {153},
    Number = {2},
    Pages = {024109},
    Month = {07},
}

@article{Karlin:1993,
  title={Metalloenzymes, structural motifs, and inorganic models},
  author={Karlin, Kenneth D},
  journal={Science},
  volume={261},
  number={5122},
  pages={701--708},
  year={1993},
  publisher={American Association for the Advancement of Science}
}

@article{Nam:2007,
  title={Dioxygen activation by metalloenzymes and models},
  author={Nam, Wonwoo},
  journal={Accounts of Chemical Research},
  volume={40},
  number={7},
  pages={465--465},
  year={2007},
  publisher={ACS Publications}
}

@article{Ribbe:2009,
  title={Molybdenum cofactors, enzymes and pathways},
  author={Schwarz, G{\"u}nter and Mendel, Ralf R and Ribbe, Markus W},
  journal={Nature},
  volume={460},
  number={7257},
  pages={839--847},
  year={2009},
  publisher={Nature Publishing Group UK London}
}

@article{Richard:1999,
  title = {Ab Initio Quantum Chemistry Using the Density Matrix Renormalization Group},
  author = {White, Steven R. and Martin, Richard L.},
  year = {1999},
  journal = {The Journal of Chemical Physics},
  volume = {110},
  number = {9},
  pages = {4127--4130},
  doi = {10.1063/1.478295}
}

@article{Chan:2011,
  title = {The {{Density Matrix Renormalization Group}} in {{Quantum Chemistry}}},
  author = {Chan, Garnet Kin-Lic and Sharma, Sandeep},
  year = {2011},
  journal = {Annual Review of Physical Chemistry},
  volume = {62},
  number = {Volume 62, 2011},
  pages = {465--481},
  doi = {10.1146/annurev-physchem-032210-103338}
}

@article{Knecht:2016,
  title = {New {{Approaches}} for Ab Initio {{Calculations}} of {{Molecules}} with {{Strong Electron Correlation}}},
  author = {Knecht, Stefan and Hedeg{\aa}rd, Erik Donovan and Keller, Sebastian and Kovyrshin, Arseny and Ma, Yingjin and Muolo, Andrea and Stein, Christopher J. and Reiher, Markus},
  year = {2016},
  journal = {CHIMIA},
  volume = {70},
  number = {4},
  pages = {244--244},
  doi = {10.2533/chimia.2016.244}
}

@article{Li:2018,
  title = {Fast Semistochastic Heat-Bath Configuration Interaction},
  author = {Li, Junhao and Otten, Matthew and Holmes, Adam A. and Sharma, Sandeep and Umrigar, C. J.},
  year = {2018},
  journal = {The Journal of Chemical Physics},
  volume = {149},
  number = {21},
  pages = {214110},
  doi = {10.1063/1.5055390}
}

@article{Smith:2017,
  title = {Cheap and {{Near Exact CASSCF}} with {{Large Active Spaces}}},
  author = {Smith, James E. T. and Mussard, Bastien and Holmes, Adam A. and Sharma, Sandeep},
  year = {2017},
  journal = {Journal of Chemical Theory and Computation},
  volume = {13},
  number = {11},
  pages = {5468--5478},
  doi = {10.1021/acs.jctc.7b00900}
}

@article{Holmes:2016,
  title = {Heat-{{Bath Configuration Interaction}}: {{An Efficient Selected Configuration Interaction Algorithm Inspired}} by {{Heat-Bath Sampling}}},
  author = {Holmes, Adam A. and Tubman, Norm M. and Umrigar, C. J.},
  year = {2016},
  journal = {Journal of Chemical Theory and Computation},
  volume = {12},
  number = {8},
  pages = {3674--3680},
  doi = {10.1021/acs.jctc.6b00407}
}

@article{Mazziotti:2011,
  title = {Large-{{Scale Semidefinite Programming}} for {{Many-Electron Quantum Mechanics}}},
  author = {Mazziotti, David A.},
  year = {2011},
  journal = {Physical Review Letters},
  volume = {106},
  number = {8},
  pages = {083001},
  doi = {10.1103/PhysRevLett.106.083001}
}

@article{Mazziotti:2012,
  title = {Two-{{Electron Reduced Density Matrix}} as the {{Basic Variable}} in {{Many-Electron Quantum Chemistry}} and {{Physics}}},
  author = {Mazziotti, David A.},
  year = {2012},
  journal = {Chemical Reviews},
  volume = {112},
  number = {1},
  pages = {244--262},
  doi = {10.1021/cr2000493}
}

@article{Mostafanejad:2019,
  title = {Combining {{Pair-Density Functional Theory}} and {{Variational Two-Electron Reduced-Density Matrix Methods}}},
  author = {Mostafanejad, Mohammad and DePrince, A. Eugene III},
  year = {2019},
  journal = {Journal of Chemical Theory and Computation},
  volume = {15},
  number = {1},
  pages = {290--302},
  doi = {10.1021/acs.jctc.8b00988}
}

@article{Mostafanejad:2020,
  title = {Global {{Hybrid Multiconfiguration Pair-Density Functional Theory}}},
  author = {Mostafanejad, Mohammad and Liebenthal, Marcus Dante and DePrince, A. Eugene III},
  year = {2020},
  journal = {Journal of Chemical Theory and Computation},
  volume = {16},
  number = {4},
  pages = {2274--2283},
  doi = {10.1021/acs.jctc.9b01178}
}

@article{Gagliardi:2017,
  title = {Multiconfiguration {{Pair-Density Functional Theory}}: {{A New Way To Treat Strongly Correlated Systems}}},
  shorttitle = {Multiconfiguration {{Pair-Density Functional Theory}}},
  author = {Gagliardi, Laura and Truhlar, Donald G. and Li Manni, Giovanni and Carlson, Rebecca K. and Hoyer, Chad E. and Bao, Junwei Lucas},
  year = {2017},
  journal = {Acc. Chem. Res.},
  volume = {50},
  number = {1},
  pages = {66--73},
  doi = {10.1021/acs.accounts.6b00471}
}

@article{Sharma:2021,
  title = {Multiconfiguration {{Pair-Density Functional Theory}}},
  author = {Sharma, Prachi and Bao, Jie J. and Truhlar, Donald G. and Gagliardi, Laura},
  year = {2021},
  journal = {Annual Review of Physical Chemistry},
  volume = {72},
  number = {Volume 72, 2021},
  pages = {541--564},
  doi = {10.1146/annurev-physchem-090419-043839}
}

@article{Zhou:2022,
  title = {Electronic Structure of Strongly Correlated Systems: Recent Developments in Multiconfiguration Pair-Density Functional Theory and Multiconfiguration Nonclassical-Energy Functional Theory},
  author = {Zhou, Chen and Hermes, Matthew R. and Wu, Dihua and Bao, Jie J. and Pandharkar, Riddhish and King, Daniel S. and Zhang, Dayou and Scott, Thais R. and Lykhin, Aleksandr O. and Gagliardi, Laura and Truhlar, Donald G.},
  year = {2022},
  journal = {Chemical Science},
  volume = {13},
  number = {26},
  pages = {7685--7706},
  doi = {10.1039/D2SC01022D}
}

@article{Bao:2025,
  title = {A Hybrid Meta On-Top Functional for Multiconfiguration Pair-Density Functional Theory},
  author = {Bao, Jie J. and Zhang, Dayou and Zhang, Shaoting and Gagliardi, Laura and Truhlar, Donald G.},
  year = {2025},
  journal = {Proceedings of the National Academy of Sciences},
  volume = {122},
  number = {1},
  pages = {e2419413121},
  doi = {10.1073/pnas.2419413121}
}

@article{Manni:2014,
  title = {Multiconfiguration {{Pair-Density Functional Theory}}},
  author = {Li Manni, Giovanni and Carlson, Rebecca K. and Luo, Sijie and Ma, Dongxia and Olsen, Jeppe and Truhlar, Donald G. and Gagliardi, Laura},
  year = {2014},
  journal = {J. Chem. Theory Comput.},
  volume = {10},
  number = {9},
  pages = {3669--3680},
  doi = {10.1021/ct500483t}
}

@article{Feng:2025,
  title = {A Cross-Entropy Corrected Hybrid Multiconfiguration Pair-Density Functional Theory for Complex Molecular Systems},
  author = {Feng, Rulin and Zhang, Igor Ying and Xu, Xin},
  year = {2025},
  journal = {Nature Communications},
  volume = {16},
  number = {1},
  pages = {235},
  doi = {10.1038/s41467-024-55524-z}
}

@article{Hennefarth:2025,
author = {Hennefarth, Matthew R. and Kim, Younghwan and Jangid, Bhavnesh and Wardzala, Jacob and Hermes, Matthew R. and Truhlar, Donald G. and Gagliardi, Laura},
title = {MC-PDFT Nuclear Gradients and L-PDFT Energies with Meta and Hybrid Meta On-Top Functionals for Ground- and Excited-State Geometry Optimization and Vertical Excitation Energies},
journal = {Journal of Chemical Theory and Computation},
volume = {21},
number = {16},
pages = {7890-7902},
year = {2025},
}

@article{Zhang:2020,
author = {Zhang, Dayou and Truhlar, Donald G.},
title = {Spin Splitting Energy of Transition Metals: A New, More Affordable Wave Function Benchmark Method and Its Use to Test Density Functional Theory},
journal = {Journal of Chemical Theory and Computation},
volume = {16},
number = {7},
pages = {4416-4428},
year = {2020},
}

@article{Siegbahn:1981,
  title = {The Complete Active Space {{SCF}} ({{CASSCF}}) Method in a {{Newton-Raphson}} Formulation with Application to the {{HNO}} Molecule},
  author = {Siegbahn, Per E. M. and Alml{\"o}f, Jan and Heiberg, Anders and Roos, Bj{\"o}rn O.},
  year = {1981},
  journal = {Journal of Chemical Physics},
  volume = {74},
  pages = {2384--2396},
  doi = {10.1063/1.441359},
}

@incollection{Roos:1987,
  title = {The {{Complete Active Space Self-Consistent Field Method}} and Its {{Applications}} in {{Electronic Structure Calculations}}},
  booktitle = {Advances in {{Chemical Physics}}},
  author = {Roos, Bj{\"o}rn O.},
  year = {1987},
  pages = {399--445},
  doi = {10.1002/9780470142943.ch7},
  isbn = {978-0-470-14294-3},
  publisher = {John Wiley \& Sons, Ltd}
}

@article{Lischka:2018,
author = {Lischka, Hans and Nachtigallová, Dana and Aquino, Ad{\'e}lia J.
A. and Szalay, P{\'e}ter G. and Plasser, Felix and Machado, Francisco B. C. and Barbatti, Mario},
title = {Multireference Approaches for Excited States of Molecules},
journal = {Chemical Reviews},
volume = {118},
number = {15},
pages = {7293-7361},
year = {2018},
}

@misc{vanvoorhis:2025,
      title={All Or Nothing: No-Downfolding Theorems For Quantum Simulation}, 
      author={Troy Van Voorhis},
      year={2025},
      eprint={2506.16534},
      archivePrefix={arXiv},
      primaryClass={quant-ph},
      url={https://arxiv.org/abs/2506.16534}, 
}

@article{Evangelisti:1987aa,
	author = {Evangelisti, S. and Daudey, J. P. and Malrieu, J. P.},
	journal = {Phys. Rev. A},
	month = {Jun},
	pages = {4930--4941},
	title = {Qualitative intruder-state problems in effective Hamiltonian theory and their solution through intermediate Hamiltonians},
	volume = {35},
	year = {1987}}

@article{Smart:2022,
author = {Smart, Scott E. and Mazziotti, David A.},
title = {Accelerated Convergence of Contracted Quantum Eigensolvers through a Quasi-Second-Order, Locally Parameterized Optimization},
journal = {Journal of Chemical Theory and Computation},
volume = {18},
number = {9},
pages = {5286-5296},
year = {2022},
}

@article{Boyn:2022aa,
	author = {Boyn, Jan-Niklas and Mazziotti, David A.},
	journal = {The Journal of Chemical Physics},
	month = {05},
	number = {19},
	pages = {194104},
	title = {Elucidating the molecular orbital dependence of the total electronic energy in multireference problems},
	volume = {156},
	year = {2022}}

@article{Boyn:2021aa,
	author = {Boyn, Jan-Niklas and Mazziotti, David A.},
	journal = {The Journal of Chemical Physics},
	month = {04},
	number = {13},
	pages = {134103},
	title = {Accurate singlet--triplet gaps in biradicals via the spin averaged anti-Hermitian contracted Schr{\"o}dinger equation},
	volume = {154},
	year = {2021}}

@article{Olivares-Amaya:2015ab,
	author = {Olivares-Amaya,Roberto and Hu,Weifeng and Nakatani,Naoki and Sharma,Sandeep and Yang,Jun and Chan,Garnet Kin-Lic},
	journal = {The Journal of Chemical Physics},
	month = {2022/12/15},
	number = {3},
	pages = {034102},
	title = {The ab-initio density matrix renormalization group in practice},
	volume = {142},
	year = {2015}}

@article{Sand:2015aa,
	author = {Sand, Andrew M. and Mazziotti, David A.},
	journal = {The Journal of Chemical Physics},
	month = {10},
	number = {13},
	pages = {134110},
	title = {Enhanced computational efficiency in the direct determination of the two-electron reduced density matrix from the anti-Hermitian contracted Schr{\"o}dinger equation with application to ground and excited states of conjugated $\pi$-systems},
	volume = {143},
	year = {2015}}

@article{Greenman:2010aa,
	author = {Greenman, Loren and Mazziotti, David A.},
	journal = {The Journal of Physical Chemistry A},
	month = {01},
	number = {1},
	pages = {583--588},
	title = {Energy Barriers of Vinylidene Carbene Reactions from the Anti-Hermitian Contracted Schr{\"o}dinger Equation},
	volume = {114},
	year = {2010}}

@article{Gidofalvi:2009aa,
	author = {Gidofalvi, Gergely and Mazziotti, David A.},
	journal = {Phys. Rev. A},
	month = {Aug},
	pages = {022507},
	title = {Direct calculation of excited-state electronic energies and two-electron reduced density matrices from the anti-Hermitian contracted Schr\"odinger equation},
	volume = {80},
	year = {2009}}

@article{Mazziotti:2007ab,
	author = {Mazziotti, David A.},
	journal = {Phys. Rev. A},
	month = {Nov},
	pages = {052502},
	title = {Multireference many-electron correlation energies from two-electron reduced density matrices computed by solving the anti-Hermitian contracted Schr\"odinger equation},
	volume = {76},
	year = {2007}}

@article{Mazziotti:2007ac,
	author = {Mazziotti, David A.},
	journal = {The Journal of Chemical Physics},
	month = {05},
	number = {18},
	pages = {184101},
	title = {Two-electron reduced density matrices from the anti-Hermitian contracted Schr{\"o}dinger equation: Enhanced energies and properties with larger basis sets},
	volume = {126},
	year = {2007}}

@article{Mazziotti:2006aa,
	author = {Mazziotti, David A.},
	journal = {Phys. Rev. Lett.},
	month = {Oct},
	pages = {143002},
	title = {Anti-Hermitian Contracted Schr\"odinger Equation: Direct Determination of the Two-Electron Reduced Density Matrices of Many-Electron Molecules},
	volume = {97},
	year = {2006}}

@article{Alcoba:2005aa,
	author = {Alcoba, D. R. and Casquero, F. J. and Tel, L. M. and P{\'e}rez-Romero, E. and Valdemoro, C.},
	journal = {International Journal of Quantum Chemistry},
	number = {5},
	pages = {620-628},
	title = {Convergence enhancement in the iterative solution of the second-order contracted Schr{\"o}dinger equation},
	volume = {102},
	year = {2005}}

@article{Nakatsuji:2002aa,
	author = {Nakatsuji, Hiroshi},
	journal = {Phys. Rev. A},
	month = {May},
	pages = {052122},
	title = {Inverse Schr\"odinger equation and the exact wave function},
	volume = {65},
	year = {2002}}

@article{Yasuda:2002aa,
	author = {Yasuda, Koji},
	journal = {Phys. Rev. A},
	month = {May},
	pages = {052121},
	title = {Uniqueness of the solution of the contracted Schr\"odinger equation},
	volume = {65},
	year = {2002}}

@article{Nakatsuji:2000aa,
	author = {Nakatsuji, Hiroshi},
	journal = {The Journal of Chemical Physics},
	month = {08},
	number = {8},
	pages = {2949-2956},
	title = {Structure of the exact wave function},
	volume = {113},
	year = {2000}}

@article{Nooijen:2000aa,
	author = {Nooijen, Marcel},
	journal = {Phys. Rev. Lett.},
	month = {Mar},
	pages = {2108--2111},
	title = {Can the Eigenstates of a Many-Body Hamiltonian Be Represented Exactly Using a General Two-Body Cluster Expansion?},
	volume = {84},
	year = {2000}}

@inbook{Valdemoro:2000aa,
	address = {Boston, MA},
	author = {Valdemoro, Carmela and Tel, L. M. and P{\'e}rez-Romero, E.},
	editor = {Cioslowski, Jerzy},
	pages = {117--137},
	publisher = {Springer US},
	title = {Critical Questions Concerning Iterative Solution of the Contracted Schr{\"o}dinger Equation},
	year = {2000}}

@article{Mazziotti:1999aa,
	author = {Mazziotti, David A.},
	journal = {Phys. Rev. A},
	month = {Dec},
	pages = {4396--4408},
	title = {Comparison of contracted Schr\"odinger and coupled-cluster theories},
	volume = {60},
	year = {1999}}

@article{Colmenero:1994aa,
	author = {Colmenero, F. and Valdemoro, C.},
	journal = {International Journal of Quantum Chemistry},
	number = {6},
	pages = {369-388},
	title = {Self-consistent approximate solution of the second-order contracted Schr{\"o}udinger equation},
	volume = {51},
	year = {1994}}

@article{Mazziotti:1998aa,
	author = {Mazziotti, David A.},
	journal = {Phys. Rev. A},
	month = {Jun},
	pages = {4219--4234},
	title = {Contracted Schr\"odinger equation: Determining quantum energies and two-particle density matrices without wave functions},
	volume = {57},
	year = {1998}}

@article{Harriman:1979aa,
	author = {Harriman, John E.},
	journal = {Phys. Rev. A},
	month = {May},
	pages = {1893--1895},
	title = {Limitation on the density-equation approach to many-electron problems},
	volume = {19},
	year = {1979}}

@article{Nakatsuji:1976aa,
	author = {Nakatsuji, Hiroshi},
	journal = {Phys. Rev. A},
	month = {Jul},
	pages = {41--50},
	title = {Equation for the direct determination of the density matrix},
	volume = {14},
	year = {1976}}

@article{Cohen:1976aa,
	author = {Cohen, Leon and Frishberg, C.},
	journal = {Phys. Rev. A},
	month = {Mar},
	pages = {927--930},
	title = {Hierarchy equations for reduced density matrices},
	volume = {13},
	year = {1976}}

@article{Colmenero:1993aa,
	author = {Colmenero, F. and P\'erez del Valle, C. and Valdemoro, C.},
	journal = {Phys. Rev. A},
	month = {Feb},
	pages = {971--978},
	title = {Approximating q-order reduced density matrices in terms of the lower-order ones. I. General relations},
	volume = {47},
	year = {1993}}

@article{Lyakh:2012aa,
	author = {Lyakh, Dmitry I. and Musia{\l}, Monika and Lotrich, Victor F. and Bartlett, Rodney J.},
	journal = {Chemical Reviews},
	month = {01},
	number = {1},
	pages = {182--243},
	title = {Multireference Nature of Chemistry: The Coupled-Cluster View},
	volume = {112},
	year = {2012}}

@article{Marian:2012aa,
	author = {Marian, Christel M.},
	journal = {WIREs Computational Molecular Science},
	month = {2025/04/07},
	number = {2},
	pages = {187--203},
	title = {Spin--orbit coupling and intersystem crossing in molecules},
	volume = {2},
	year = {2012}}

@article{Curchod:2018aa,
    Author = {Curchod, Basile F. E. and Martínez, Todd J.},
    Title = {Ab Initio Nonadiabatic Quantum Molecular Dynamics},
    Journal = {Chemical Reviews},
    Year = {2018},
    Volume = {118},
    Number = {7},
    Pages = {3305--3336},
    Month = {04},
}

@article{Singh:2020aa,
    Author = {Singh, Saurabh Kumar and Cramer, Christopher J. and Gagliardi, Laura},
    Title = {Correlating Electronic Structure and Magnetic Anisotropy in Actinide Complexes [An(COT)2], AnIII/IV = U, Np, and Pu},
    Journal = {Inorganic Chemistry},
    Year = {2020},
    Volume = {59},
    Number = {10},
    Pages = {6815-6825},
}

@article{Singh:2018aa,
    Author = {Singh, Saurabh Kumar and Atanasov, Mihail and Neese, Frank},
    Title = {Challenges in Multireference Perturbation Theory for the Calculations of the g-Tensor of First-Row Transition-Metal Complexes},
    Journal = {Journal of Chemical Theory and Computation},
    Year = {2018},
    Volume = {14},
    Number = {9},
    Pages = {4662--4677},
    Month = {09},
}

@article{Averkiev:2011aa,
    Author = {Averkiev, Boris B. and Mantina, Manjeera and Valero, Rosendo and Infante, Ivan and Kovacs, Attila and Truhlar, Donald G. and Gagliardi, Laura},
    Title = {How accurate are electronic structure methods for actinoid chemistry?},
    Journal = {Theoretical Chemistry Accounts},
    Year = {2011},
    Volume = {129},
    Number = {3},
    Pages = {657--666},
}

@article{Foley:2021aa,
    author = {Foley, Jonathan J., IV and Rothman, Adam E. and Mazziotti, David A.},
    title = {Strongly correlated mechanisms of a photoexcited radical reaction from the anti-Hermitian contracted Schrödinger equation},
    journal = {The Journal of Chemical Physics},
    volume = {134},
    number = {3},
    pages = {034111},
    year = {2011},
    month = {01},
    issn = {0021-9606},
}

@article{Snyder:2010aa,
    author = {Snyder, James W., Jr. and Rothman, Adam E. and Foley, Jonathan J., IV and Mazziotti, David A.},
    title = {Conical intersections in triplet excited states of methylene from the anti-Hermitian contracted Schrödinger equation},
    journal = {The Journal of Chemical Physics},
    volume = {132},
    number = {15},
    pages = {154109},
    year = {2010},
    month = {04},
    issn = {0021-9606},
}

@article{Snyder:2011aa,
author = {Snyder, James W. Jr. and Mazziotti, David A.},
title = {Conical Intersection of the Ground and First Excited States of Water: Energies and Reduced Density Matrices from the Anti-Hermitian Contracted Schrödinger Equation},
journal = {The Journal of Physical Chemistry A},
volume = {115},
number = {48},
pages = {14120-14126},
year = {2011},
}

@article{Foley:2009aa,
    author = {Foley, Jonathan J., IV and Rothman, Adam E. and Mazziotti, David A.},
    title = {Activation energies of sigmatropic shifts in propene and acetone enolate from the anti-Hermitian contracted Schrödinger equation},
    journal = {The Journal of Chemical Physics},
    volume = {130},
    number = {18},
    pages = {184112},
    year = {2009},
    month = {05},
    issn = {0021-9606},
}

@article{Greenman:2011aa,
    author = {Greenman, Loren and Mazziotti, David A.},
    title = {Balancing single- and multi-reference correlation in the chemiluminescent reaction of dioxetanone using the anti-Hermitian contracted Schrödinger equation},
    journal = {The Journal of Chemical Physics},
    volume = {134},
    number = {17},
    pages = {174110},
    year = {2011},
    month = {05},
    issn = {0021-9606},
}

@article{Mazziotti:2008aa,
author = {Mazziotti, David A.},
title = {Energy Barriers in the Conversion of Bicyclobutane to gauche-1,3-Butadiene from the Anti-Hermitian Contracted Schrödinger Equation},
journal = {The Journal of Physical Chemistry A},
volume = {112},
number = {51},
pages = {13684-13690},
year = {2008},
}

@book{szabo:1989aa,
  address = {Mineola},
  author = {Szabo, Attila and Ostlund, Neil S.},
  edition = {First},
  publisher = {Dover Publications, Inc.},
  title = {Modern Quantum Chemistry: Introduction to Advanced Electronic
                  Structure Theory},
  year = 1996
}

@article{Hehre:1972aa,
    author = {Hehre, W. J. and Ditchfield, R. and Pople, J. A.},
    title = {Self—Consistent Molecular Orbital Methods. XII. Further Extensions of Gaussian—Type Basis Sets for Use in Molecular Orbital Studies of Organic Molecules},
    journal = {The Journal of Chemical Physics},
    volume = {56},
    number = {5},
    pages = {2257-2261},
    year = {1972},
    month = {03},
    issn = {0021-9606},
}

@Article{Hariharan:1973aa,
author="Hariharan, P. C.
and Pople, J. A.",
title="The influence of polarization functions on molecular orbital hydrogenation energies",
journal="Theoretica chimica acta",
year="1973",
month="Sep",
day="01",
volume="28",
number="3",
pages="213--222",
issn="1432-2234",
}

@article{Hariharan:1974aa,
author = {P.C. Hariharan and J.A. Pople},
title = {Accuracy of AH n equilibrium geometries by single determinant molecular orbital theory},
journal = {Molecular Physics},
volume = {27},
number = {1},
pages = {209--214},
year = {1974},
publisher = {Taylor \& Francis},

}

@article{Ditchfield:1971aa,
    author = {Ditchfield, R. and Hehre, W. J. and Pople, J. A.},
    title = {Self‐Consistent Molecular‐Orbital Methods. IX. An Extended Gaussian‐Type Basis for Molecular‐Orbital Studies of Organic Molecules},
    journal = {The Journal of Chemical Physics},
    volume = {54},
    number = {2},
    pages = {724-728},
    year = {1971},
    month = {01},
    issn = {0021-9606},
}

@article{Dunning:1989aa,
    author = {Dunning, Thom H., Jr.},
    title = {Gaussian basis sets for use in correlated molecular calculations. I. The atoms boron through neon and hydrogen},
    journal = {The Journal of Chemical Physics},
    volume = {90},
    number = {2},
    pages = {1007-1023},
    year = {1989},
    month = {01},
    issn = {0021-9606},
}

@article{Kendall:1992aa,
    author = {Kendall, Rick A. and Dunning, Thom H., Jr. and Harrison, Robert J.},
    title = {Electron affinities of the first‐row atoms revisited. Systematic basis sets and wave functions},
    journal = {The Journal of Chemical Physics},
    volume = {96},
    number = {9},
    pages = {6796-6806},
    year = {1992},
    month = {05},
    issn = {0021-9606},
}

@Article{Weigend:2005aa,
author ="Weigend, Florian and Ahlrichs, Reinhart",
title  ="Balanced basis sets of split valence{,} triple zeta valence and quadruple zeta valence quality for H to Rn: Design and assessment of accuracy",
journal  ="Phys. Chem. Chem. Phys.",
year  ="2005",
volume  ="7",
issue  ="18",
pages  ="3297-3305",
publisher  ="The Royal Society of Chemistry",
}

@Article{Weigend:2006aa,
author ="Weigend, Florian",
title  ="Accurate Coulomb-fitting basis sets for H to Rn",
journal  ="Phys. Chem. Chem. Phys.",
year  ="2006",
volume  ="8",
issue  ="9",
pages  ="1057-1065",
publisher  ="The Royal Society of Chemistry",
}

@article{Zhai:2021aa,
    author = {Zhai, Huanchen and Chan, Garnet Kin-Lic},
    title = {Low communication high performance ab initio density matrix renormalization group algorithms},
    journal = {The Journal of Chemical Physics},
    volume = {154},
    number = {22},
    pages = {224116},
    year = {2021},
    month = {06},
    issn = {0021-9606},
}

@article{Zhai:2023aa,
    author = {Zhai, Huanchen and Larsson, Henrik R. and Lee, Seunghoon and Cui, Zhi-Hao and Zhu, Tianyu and Sun, Chong and Peng, Linqing and Peng, Ruojing and Liao, Ke and Tölle, Johannes and Yang, Junjie and Li, Shuoxue and Chan, Garnet Kin-Lic},
    title = {Block2: A comprehensive open source framework to develop and apply state-of-the-art DMRG algorithms in electronic structure and beyond},
    journal = {The Journal of Chemical Physics},
    volume = {159},
    number = {23},
    pages = {234801},
    year = {2023},
    month = {12},
    issn = {0021-9606},
}

@article{Angeli:2001aa,
    author = {Angeli, C. and Cimiraglia, R. and Evangelisti, S. and Leininger, T. and Malrieu, J.-P.},
    title = {Introduction of n-electron valence states for multireference perturbation theory},
    journal = {The Journal of Chemical Physics},
    volume = {114},
    number = {23},
    pages = {10252-10264},
    year = {2001},
    month = {06},
    issn = {0021-9606},
}

@article{Angeli:2001ab,
title = {N-electron valence state perturbation theory: a fast implementation of the strongly contracted variant},
journal = {Chemical Physics Letters},
volume = {350},
number = {3},
pages = {297-305},
year = {2001},
issn = {0009-2614},
}

@article{Mazziotti:2004aa,
  title = {Exactness of wave functions from two-body exponential transformations in many-body quantum theory},
  author = {Mazziotti, David A.},
  journal = {Phys. Rev. A},
  volume = {69},
  issue = {1},
  pages = {012507},
  numpages = {11},
  year = {2004},
  month = {Jan},
  publisher = {American Physical Society},
}

@article{Mazziotti:2007ad,
  title = {Anti-Hermitian part of the contracted Schr\"odinger equation for the direct calculation of two-electron reduced density matrices},
  author = {Mazziotti, David A.},
  journal = {Phys. Rev. A},
  volume = {75},
  issue = {2},
  pages = {022505},
  numpages = {12},
  year = {2007},
  month = {Feb},
  publisher = {American Physical Society},
}

@article{Phillips:2011aa,
    author = {Phillips, Jordan J. and Peralta, Juan E. and Janesko, Benjamin G.},
    title = {Magnetic exchange couplings evaluated with Rung 3.5 density functionals},
    journal = {The Journal of Chemical Physics},
    volume = {134},
    number = {21},
    pages = {214101},
    year = {2011},
    month = {06},
    issn = {0021-9606},
}

@article{NIST,
    author = {Kramida, A. and Ralchenko, Yu. and Reader, J. and NIST ASD Team},
    title = {NIST Atomic Spectra Database (version 5.12)},
    year = 2024,
    url = {https://physics.nist.gov/asd},
    doi = {https://doi.org/10.18434/T4W30F},
}

@article{Schlimgen:2017aa,
	author = {Schlimgen, Anthony W. and Mazziotti, David A.},
	journal = {The Journal of Physical Chemistry A},
	month = {12},
	number = {48},
	pages = {9377--9384},
	title = {Static and Dynamic Electron Correlation in the Ligand Noninnocent Oxidation of Nickel Dithiolates},
	volume = {121},
	year = {2017}}

@article{ACSE_github,
    author = {Gibney, D. and Schlimgen, A. and Boyn, J.-N.},
    title = {{BoynGroup/ACSE}},
    year = 2026,
    url = {https://github.com/BoynGroup/ACSE},
    doi = {https://doi.org/10.5281/zenodo.19390063},
}

@article{Foley:2011aa,
	author = {Foley, Jonathan J., IV and Rothman, Adam E. and Mazziotti, David A.},
	journal = {The Journal of Chemical Physics},
	month = {01},
	number = {3},
	pages = {034111},
	title = {Strongly correlated mechanisms of a photoexcited radical reaction from the anti-Hermitian contracted Schr{\"o}dinger equation},
	volume = {134},
	year = {2011}}

@article{DePrince:2007aa,
	author = {DePrince, A. Eugene, III and Mazziotti, David A.},
	journal = {The Journal of Chemical Physics},
	month = {3/31/2026},
	number = {10},
	pages = {104104},
	title = {Cumulant reconstruction of the three-electron reduced density matrix in the anti-Hermitian contracted Schr{\"o}dinger equation},
	volume = {127},
	year = {2007}}

\clearpage
\appendix
\section{Spin block normalizations}

\begin{equation}
    \begin{aligned}
    Tr({}^2D_{\alpha\alpha})&=\frac{N_\alpha(N_\alpha-1)}{2}\\
    Tr({}^2D_{\alpha\alpha})&=\frac{N_\alpha(N_\beta)}{2}\\
    Tr({}^2D_{\alpha\alpha})&=\frac{N_\beta(N_\beta-1)}{2}\\
    Tr({}^3D_{\alpha\alpha\alpha})&=\frac{N_\alpha(N_\alpha-1)(N_\alpha-2)}{3!}\\
    Tr({}^3D_{\alpha\alpha\beta})&=\frac{N_\alpha(N_\alpha-1)(N_\beta)}{3!}\\
    Tr({}^3D_{\alpha\beta\beta})&=\frac{N_\alpha(N_\beta)(N_\beta-1)}{3!}\\
    \end{aligned}
\end{equation}

\section{Obtaining the residual expression in terms of the 2- and 3-RDMs}
Using the anti-commutation relations of fermionic operators, we write the normal-ordered residual from Eq. \ref{eq:residual_2quant},
\begin{equation}
    \begin{aligned}
        R^{ij}_{kl} &= \langle \psi|[a^\dagger_ia^\dagger_ja_la_k,H]|\psi\rangle\\
        &=\langle \psi|[a^\dagger_ia^\dagger_ja_la_k,{}^2K^{pq}_{sr}a^\dagger_pa^\dagger_qa_ra_s]|\psi\rangle\\
        &=\sum_{pqrs}{}^2K^{pq}_{sr}\langle\psi|a^\dagger_ia^\dagger_ja_ra_s\delta_{kq}\delta_{lp}\\&+a^\dagger_ia^\dagger_ja^\dagger_qa_la_ra_s\delta_{kp}-a^\dagger_ia^\dagger_ja_ra_s\delta_{kp}\delta_{lq}
        \\&-a^\dagger_ia^\dagger_ja^\dagger_qa_ka_ra_s\delta_{lp}-a^\dagger_ia^\dagger_ja^\dagger_pa_la_ra_s\delta_{kq}\\&+a^\dagger_ia^\dagger_ja^\dagger_pa_ka_ra_s\delta_{lq}+a^\dagger_pa^\dagger_qa_ka_l\delta_{ir}\delta_{js}\\&-a^\dagger_ja^\dagger_pa^\dagger_qa_ka_la_s\delta_{ir}-a^\dagger_pa^\dagger_qa_ka_l\delta_{is}\delta_{jr}
        \\&+a^\dagger_ja^\dagger_pa^\dagger_qa_ka_la_r\delta_{is}+a^\dagger_ia^\dagger_pa^\dagger_qa_ka_la_s\delta_{jr}\\&-a^\dagger_ia^\dagger_pa^\dagger_qa_ka_la_r\delta_{js}|\psi\rangle.
    \end{aligned}
\end{equation}
We write in terms of the density matrices,
\begin{equation}
    \begin{aligned}
        R^{ij}_{kl} &= \sum_{pqrs}{}^2K^{pq}_{sr}[3~{}^3D^{ijq}_{srl}\delta_{kp}-{}^2D^{ij}_{sr}\delta_{kp}\delta_{lq}\\&+{}^2D^{ij}_{sr}\delta_{kq}\delta_{lp}-3~{}^3D^{ijq}_{srk}\delta_{lp}-3~{}^3D^{ijp}_{srl}\delta_{kq}\\&+3~{}^3D^{ijp}_{srk}\delta_{lq}+{}^2D^{pq}_{lk}\delta_{ir}\delta_{js}-3~{}^3D^{jpq}_{slk}\delta_{ir}\\&-{}^2D^{pq}_{lk}\delta_{is}\delta_{jr}+3~{}^3D^{jpq}_{rlk}\delta_{is}+3~{}^3D^{ipq}_{slk}\delta_{jr}\\&-3~{}^3D^{ipq}_{rlk}\delta_{js}],
    \end{aligned}
\end{equation}
and resolve delta functions,
\begin{equation}
    \begin{aligned}
        R^{ij}_{kl} &= \sum_{pqrs}[3~{}^2K^{kq}_{sr}{}^3D^{ijq}_{srl}-{}^2K^{kl}_{sr}{}^2D^{ij}_{sr}+3~{}^2K^{lk}_{sr}{}^2D^{ij}_{sr}\\
        &-3~{}^2K^{lq}_{sr}{}^3D^{ijq}_{srk}-3~{}^2K^{pk}_{sr}{}^3D^{ijp}_{srl}+3~{}^2K^{pl}_{sr}{}^3D^{ijp}_{srk}\\
        &+3~{}^2K^{pq}_{ji}{}^2D^{pq}_{lk}-3~{}^2K^{pq}_{si}{}^3D^{jpq}_{slk}-{}^2K^{pq}_{ij}{}^2D^{pq}_{lk}\\
        &+3~{}^2K^{pq}_{ir}{}^3D^{jpq}_{rlk}+3~{}^2K^{pq}_{sj}{}^3D^{ipq}_{slk}-3~{}^2K^{pq}_{jr}{}^3D^{ipq}_{rlk}].
    \end{aligned}
\end{equation}
By rewriting the summations we eliminate $s$,
\begin{equation}
    \begin{aligned}
        R^{ij}_{kl} &= \sum_{pqr}[{}^2K^{lk}_{qp}{}^2D^{ij}_{qp}+3~{}^2K^{kp}_{rq}{}^3D^{ijp}_{rql}-{}^2K^{kl}_{qp}{}^2D^{ij}_{qp}\\
        &-3~{}^2K^{lp}_{rq}{}^3D^{ijp}_{rqk}-3~{}^2K^{pk}_{rq}{}^3D^{ijp}_{rql}+3~{}^2K^{pl}_{rq}{}^3D^{ijp}_{rqk}\\
        &+{}^2K^{pq}_{ji}{}^2D^{pq}_{lk}-3~{}^2K^{pq}_{ri}{}^3D^{jpq}_{rlk}-{}^2K^{pq}_{ij}{}^2D^{pq}_{lk}\\
        &+3~{}^2K^{pq}_{ir}{}^3D^{jpq}_{rlk}+3~{}^2K^{pq}_{rj}{}^3D^{ipq}_{rlk}-3~{}^2K^{pq}_{jr}{}^3D^{ipq}_{rlk}].
    \end{aligned}
\end{equation}
Using the Hermiticity of $K$ and ${}^2D$,
\begin{equation}
    \begin{aligned}
        R^{ij}_{kl} &= \sum_{pqr}2~{}^2K^{pq}_{ji}{}^2D^{pq}_{lk}+6~{}^2K^{kp}_{rq}{}^3D^{ijp}_{rql}-6~{}^2K^{lp}_{rq}{}^3D^{ijp}_{rqk}\\
        &-2~{}^2K^{kl}_{qp}{}^2D^{ij}_{qp}-6~{}^2K^{pq}_{ri}{}^3D^{jpq}_{rlk}+6~{}^2K^{pq}_{rj}{}^3D^{ipq}_{rlk},
    \end{aligned}
\end{equation}
and further using the $K$ symmetry, $K^{kl}_{pq} = K^{lk}_{qp}$, and the ${}^2D$ antisymmetry, ${}^2D^{kl}_{qp} = -{}^2D^{kl}_{pq}$, we find,
\begin{equation}
\begin{aligned}
    R^{ij}_{kl} &= \sum_{pqr} 2K^{ij}_{qp}~{}^2D^{kl}_{qp}+6K^{kp}_{rq}~{}^3D^{ijp}_{rql}-6K^{lp}_{rq}~{}^3D^{ijp}_{rqk} \\ 
    &-2K^{kl}_{qp}~{}^2D^{ij}_{qp} -6K^{pq}_{ri}~{}^3D^{jpq}_{rlk}+6K^{pq}_{rj}~{}^3D^{ipq}_{rlk}.
\end{aligned}
\label{eq:appendix_residual}
\end{equation}
We note that when computing the update $U$ from the residual, the anti-hermiticity of $R$, rather than the hermiticity of $^2D$, induces a negative sign when used in these relations.

The purely two-body terms,
\begin{equation}
\begin{aligned}
A^{ij}_{kl}&=-\sum_{pqr} 2K^{kl}_{qp}~{}^2D^{ij}_{qp}\\
B^{ij}_{kl}&=\sum_{pqr} 2K^{ij}_{qp}~{}^2D^{kl}_{qp}
\end{aligned}
\end{equation}
are related by transposition of indices, $A^{ij}_{kl} = B^{kl}_{ij}$, and is computed with one contraction and one transposed sum. A similar reduction is achieved with four other terms in Eq. \ref{eq:appendix_residual}, resulting in halving the number of required contractions. Explicitly,
\begin{equation}
\begin{aligned}
    A^{ij}_{kl} &= \sum_{pqr} 2K^{kl}_{qp}~{}^2D^{ij}_{qp}\\
    B^{ij}_{kl} &= \sum_{pqr}6K^{kp}_{rq}~{}^3D^{ijp}_{rql}\\
    C^{ij}_{kl} &= \sum_{pqr}-6K^{pq}_{ri}~{}^3D^{jpq}_{rlk}\\
    R^{ij}_{kl} &= A^{ij}_{kl} - A^{lk}_{ji} + B^{ij}_{kl} - B^{ij}_{lk} + C^{ij}_{kl} - C^{ji}_{kl},
\end{aligned}
\end{equation}
which is further simplified.
\begin{equation}
\begin{aligned}
    A^{ij}_{kl} &= \sum_{pqr} 2K^{kl}_{qp}~{}^2D^{ij}_{qp}\\
    B^{ij}_{kl} &= \sum_{pqr}6K^{kp}_{rq}~{}^3D^{ijp}_{rql}\\
    C^{ij}_{kl} &= B^{ij}_{kl} - B^{ij}_{lk}\\
    R^{ij}_{kl} &= A^{ij}_{kl} - A^{lk}_{ji} + C^{ij}_{kl} + C^{lk}_{ij}.
\end{aligned}
\end{equation}
Here, with the 3-RDM update construction, the final $C$ term is instead $-C^{lk}_{ij}$.
\section{Cumulant decomposition of the 3-RDM}
The 3-RDM may be rewritten in terms of its cumulant expansion,
\begin{equation}
    ^3D = (3
    {^2\Delta} + ({^1D} \wedge  {^1D})) \wedge {^1D} + {^3\Delta}.
\end{equation}
Expanding the first wedge product yields,
\begin{equation}
    ^3D^{ijp}_{rql} = (3
    {^2\Delta}^{ij}_{rq} + \frac{1}{2}({^1D^i_r}{^1D^j_q} - {^1D^j_r}{^1D^i_q})) \wedge {^1D}^p_l + {^3\Delta}^{ijp}_{rql}.
\end{equation}
For concision we define $M_{rq}^{ij}= 3{^2\Delta}^{ij}_{rq} + \frac{1}{2}({^1D^i_r}{^1D^j_q} - {^1D^j_r}{^1D^i_q})$, and expanding further,
\begin{equation}
\begin{aligned}
    ^3D^{ijp}_{rql} &= M^{ij}_{rq} \wedge {^1D}^p_l + {^3\Delta}^{ijp}_{rql}\\
    &= \frac{1}{36}(M^{ij}_{rq}{^1D}^p_l - M^{ij}_{lq}{^1D}^p_r - M^{ij}_{rl}{^1D}^p_q - M^{ij}_{qr}{^1D}^p_l \\&+ M^{ij}_{ql}{^1D}^p_r + M^{ij}_{lr}{^1D}^p_q
    - M^{ji}_{rq}{^1D}^p_l + M^{ji}_{lq}{^1D}^p_r \\&+ M^{ji}_{rl}{^1D}^p_q + M^{ji}_{qr}{^1D}^p_l - M^{ji}_{ql}{^1D}^p_r - M^{ji}_{lr}{^1D}^p_q \\
    &- M^{ip}_{rq}{^1D}^j_l + M^{ip}_{lq}{^1D}^j_r + M^{ip}_{rl}{^1D}^j_q + M^{ip}_{qr}{^1D}^j_l \\&- M^{ip}_{ql}{^1D}^j_r - M^{ip}_{lr}{^1D}^j_q
    + M^{jp}_{rq}{^1D}^i_l - M^{jp}_{lq}{^1D}^i_r \\&- M^{jp}_{rl}{^1D}^i_q - M^{jp}_{qr}{^1D}^i_l + M^{jp}_{ql}{^1D}^i_r + M^{jp}_{lr}{^1D}^i_q\\
    &- M^{pj}_{rq}{^1D}^i_l + M^{pj}_{lq}{^1D}^i_r + M^{pj}_{rl}{^1D}^i_q + M^{pj}_{qr}{^1D}^i_l \\&- M^{pj}_{ql}{^1D}^i_r - M^{pj}_{lr}{^1D}^i_q
    + M^{pi}_{rq}{^1D}^j_l - M^{pi}_{lq}{^1D}^j_r \\&- M^{pi}_{rl}{^1D}^j_q - M^{pi}_{qr}{^1D}^j_l + M^{pi}_{ql}{^1D}^j_r + M^{pi}_{lr}{^1D}^j_q) \\&+ {^3\Delta}^{ijp}_{rql}.\\
\end{aligned}
\end{equation}
Noting that M is antisymmetric in its indices,
\begin{equation}
    {}^2M^{ij}_{lk} = -{}^2M^{ji}_{lk} = -{}^2M^{ij}_{kl} = {}^2M^{ji}_{kl},
\end{equation}
we simplify,
\begin{equation}
\begin{aligned}
        ^3D^{ijp}_{rql} &= \frac{1}{9}(M^{ij}_{rq}{^1D}^p_l - M^{ij}_{lq}{^1D}^p_r - M^{ij}_{rl}{^1D}^p_q\\ 
    &- M^{ip}_{rq}{^1D}^j_l + M^{ip}_{lq}{^1D}^j_r + M^{ip}_{rl}{^1D}^j_q\\
    &+ M^{jp}_{rq}{^1D}^i_l - M^{jp}_{lq}{^1D}^i_r - M^{jp}_{rl}{^1D}^i_q) + {^3\Delta}^{ijp}_{rql}.\\
\end{aligned}
\label{eq:3rdm_recon}
\end{equation}
All ${}^2M{}^1D$ terms are related by transposition, so only a single contraction and 8 transpositions are required to reconstruct the 3-RDM. The residual can be calculated without explicitly storing the 3-RDM by inserting Eq. \ref{eq:3rdm_recon} into Eq. \ref{eq:residual}, and neglecting or approximating the 3-cumulant $^3\Delta$. 
\section{3-RDM reconstructions}
\subsection{Valdemoro}
Here we show the explicit expansion of the first 3-RDM term in Eq. \ref{eq:residual} using the Valdemoro reconstruction,
\begin{equation}
\begin{aligned}
\sum_{pqr} K^{kp}_{rq}~{}^3D^{ijp}_{rql} &= \frac{1}{9}\sum_{pqr} K^{kp}_{rq}~(M^{ij}_{rq}{^1D}^p_l - M^{ij}_{lq}{^1D}^p_r \\&- M^{ij}_{rl}{^1D}^p_q- M^{ip}_{rq}{^1D}^j_l + M^{ip}_{lq}{^1D}^j_r \\&+ M^{ip}_{rl}{^1D}^j_q+ M^{jp}_{rq}{^1D}^i_l - M^{jp}_{lq}{^1D}^i_r \\&- M^{jp}_{rl}{^1D}^i_q).
\end{aligned}
\label{eq:app_x}
\end{equation}
This is simplified by recognizing that some of the above terms are related by transpositions. 
\begin{equation}
\begin{aligned}
&\sum_{pqr} K^{kp}_{rq}~{}^3D^{ijp}_{rql} \\&= \frac{1}{9}\sum_{pqr} K^{kp}_{rq}M^{ij}_{rq}{^1D}^p_l - (K^{kp}_{rq}-K^{kp}_{qr})M^{ij}_{lq}{^1D}^p_r\\ 
    &-K^{kp}_{rq} M^{ip}_{rq}{^1D}^j_l + (K^{kp}_{rq}-K^{kp}_{qr})M^{ip}_{lq}{^1D}^j_r\\
    &+ K^{kp}_{rq}M^{jp}_{rq}{^1D}^i_l - (K^{kp}_{rq}-K^{kp}_{qr})M^{jp}_{lq}{^1D}^i_r,
\end{aligned}
\end{equation}
and further simplified with intermediate matrices,
\begin{equation}
    \begin{aligned}
        E^{ij}_{kl} = \sum_{pqr} (K^{kp}_{rq}-K^{kp}_{qr})M^{ip}_{lq}{^1D}^j_r -K^{kp}_{rq} M^{ip}_{rq}{^1D}^j_l,\\
    \end{aligned}
\end{equation}
resulting in,
\begin{equation}
\begin{aligned}
\sum_{pqr} K^{kp}_{rq}~{}^3D^{ijp}_{rql} &= \frac{1}{9}\sum_{pqr} K^{kp}_{rq}M^{ij}_{rq}{^1D}^p_l \\&- (K^{kp}_{rq}-K^{kp}_{qr})M^{ij}_{lq}{^1D}^p_r+E^{ij}_{kl}- E^{ji}_{kl}.
\end{aligned}
\end{equation}
\subsection{Nakatsuji-Yasuda}
The NY approximation to the 3-cumulant is,
\begin{equation}
    \begin{aligned}
        {}^3\Delta^{ijp}_{rql} = \frac{1}{6}\sum_{a}\sigma_{a}\hat{A}({}^2\Delta^{ia}_{rq}{}^2\Delta^{jp}_{al})
    \end{aligned}
\end{equation}
where $\sigma_a$ is 1 if $a$ is an occupied orbital and -1 if it is an unoccupied orbital in the Hartree-Fock reference. $\hat{A}$ is the antisymmetry operator which here permutes all indices except $a$. Explicitly, the resulting equation is,
\begin{equation}
    \begin{aligned}
        {}^3\Delta^{ijp}_{rql} &= \frac{1}{6}\sum_{a}\sigma_a({}^2\Delta^{ia}_{rq}\Delta^{jp}_{al}-{}^2\Delta^{ia}_{rq}\Delta^{pj}_{al}-{}^2\Delta^{ja}_{rq}\Delta^{ip}_{al}\\&-{}^2\Delta^{pa}_{rq}\Delta^{ji}_{al}+{}^2\Delta^{pa}_{rq}\Delta^{ij}_{al}+{}^2\Delta^{ja}_{rq}\Delta^{pi}_{al}
        -{}^2\Delta^{ia}_{qr}\Delta^{jp}_{al}\\&+{}^2\Delta^{ia}_{qr}\Delta^{pj}_{al}+{}^2\Delta^{ja}_{qr}\Delta^{ip}_{al}+{}^2\Delta^{pa}_{qr}\Delta^{ji}_{al}-{}^2\Delta^{pa}_{qr}\Delta^{ij}_{al}\\&-{}^2\Delta^{ja}_{qr}\Delta^{pi}_{al}
        -{}^2\Delta^{ia}_{lq}\Delta^{jp}_{ar}+{}^2\Delta^{ia}_{lq}\Delta^{pj}_{ar}+{}^2\Delta^{ja}_{lq}\Delta^{ip}_{ar}\\&+{}^2\Delta^{pa}_{lq}\Delta^{ji}_{ar}-{}^2\Delta^{pa}_{lq}\Delta^{ij}_{ar}-{}^2\Delta^{ja}_{lq}\Delta^{pi}_{ar}
        -{}^2\Delta^{ia}_{rl}\Delta^{jp}_{aq}\\&+{}^2\Delta^{ia}_{rl}\Delta^{pj}_{aq}+{}^2\Delta^{ja}_{rl}\Delta^{ip}_{aq}+{}^2\Delta^{pa}_{rl}\Delta^{ji}_{aq}-{}^2\Delta^{pa}_{rl}\Delta^{ij}_{aq}\\&-{}^2\Delta^{ja}_{rl}\Delta^{pi}_{aq}
        +{}^2\Delta^{ia}_{ql}\Delta^{jp}_{ar}-{}^2\Delta^{ia}_{ql}\Delta^{pj}_{ar}-{}^2\Delta^{ja}_{ql}\Delta^{ip}_{ar}\\&-{}^2\Delta^{pa}_{ql}\Delta^{ji}_{ar}+{}^2\Delta^{pa}_{ql}\Delta^{ij}_{ar}+{}^2\Delta^{ja}_{ql}\Delta^{pi}_{ar}
        +{}^2\Delta^{ia}_{lr}\Delta^{jp}_{aq}\\&-{}^2\Delta^{ia}_{lr}\Delta^{pj}_{aq}-{}^2\Delta^{ja}_{lr}\Delta^{ip}_{aq}-{}^2\Delta^{pa}_{lr}\Delta^{ji}_{aq}+{}^2\Delta^{pa}_{lr}\Delta^{ij}_{aq}\\&+{}^2\Delta^{ja}_{lr}\Delta^{pi}_{aq}),
    \end{aligned}
\end{equation}
and simplified to,
\begin{equation}
    \begin{aligned}
        {}^3\Delta^{ijp}_{rql} = \frac{1}{6}&\sum_{a}\sigma_a(4~{}^2\Delta^{ia}_{rq}\Delta^{jp}_{al}-4~{}^2\Delta^{ja}_{rq}\Delta^{ip}_{al}-4~{}^2\Delta^{pa}_{rq}\Delta^{ji}_{al}\\
        &-4~{}^2\Delta^{ia}_{lq}\Delta^{jp}_{ar}+4~{}^2\Delta^{ja}_{lq}\Delta^{ip}_{ar}+4~{}^2\Delta^{pa}_{lq}\Delta^{ji}_{ar}\\
        &-4~{}^2\Delta^{ia}_{rl}\Delta^{jp}_{aq}+4~{}^2\Delta^{ja}_{rl}\Delta^{ip}_{aq}+4~{}^2\Delta^{pa}_{rl}\Delta^{ji}_{aq}).\\
    \end{aligned}
\end{equation}
Here again there is only one unique term with the rest being related through transposition. The NY contribution to the residual is,
\begin{equation}
    \begin{aligned}
                R^{ij}_{kl} = \frac{1}{6}&\sum_{pqra}\sigma_aK^{kp}_{rq}(4~{}^2\Delta^{ia}_{rq}\Delta^{jp}_{al}-4~{}^2\Delta^{ja}_{rq}\Delta^{ip}_{al}\\&-4~{}^2\Delta^{pa}_{rq}\Delta^{ji}_{al}
        -4~{}^2\Delta^{ia}_{lq}\Delta^{jp}_{ar}+4~{}^2\Delta^{ja}_{lq}\Delta^{ip}_{ar}\\&+4~{}^2\Delta^{pa}_{lq}\Delta^{ji}_{ar}
        -4~{}^2\Delta^{ia}_{rl}\Delta^{jp}_{aq}+4~{}^2\Delta^{ja}_{rl}\Delta^{ip}_{aq}\\&+4~{}^2\Delta^{pa}_{rl}\Delta^{ji}_{aq}),\\
    \end{aligned}
\end{equation}
and simplified to,
\begin{equation}
    \begin{aligned}
                A^{ij}_{kl}&=\sum_{pqra}K^{kp}_{rq}\sigma_a{}^2\Delta^{ia}_{rq}{}^2\Delta^{jp}_{al}\\&~~~~~~~~~~-(K^{kp}_{rq}-K^{kp}_{qr})\sigma_a{}^2\Delta^{ia}_{lq}{}^2\Delta^{jp}_{ar}\\
                R^{ij}_{kl} &= \frac{1}{6}\sum_{pqra}K^{kp}_{rq}\sigma_a{}^2\Delta^{pa}_{rq}{}^2\Delta^{ji}_{al}\\&~~~~~~~~~-(K^{kp}_{rq}-K^{kp}_{qr})\sigma_a{}^2\Delta^{pa}_{lq}{}^2\Delta^{ji}_{ar}\\
                &~~~~~~~~~+ A^{ij}_{kl} - A^{ji}_{kl}.
    \end{aligned}
\end{equation}

\section{Ethylene Geometry}
Ethylene's geometry for the S$_0\xrightarrow{}$S$_1$ transition is defined using a Z-matrix as
\begin{lstlisting}
HC = 1.086
CC = 1.339
A1 = 117.6
D1 = 180
D2 = 0
C
C 1 CC
H 1 HC 2 A1
H 1 HC 2 A1 3 D1
H 2 HC 1 A1 3 D1
H 2 HC 1 A1 3 D2
\end{lstlisting}
while the Z-matrix parameters for the dihedral angle potential energy surface are modified with the values below,
\begin{lstlisting}
HC = 1.08
CC = 1.32
A1 = 121.6
D1 = 180,
\end{lstlisting}
and D2 is the dihedral angle to be scanned.
\end{document}